
\def\IR{{\hbox{{\rm I}\kern-.2em\hbox{\rm R}}}}
\def\IB{{\hbox{{\rm I}\kern-.2em\hbox{\rm B}}}}
\def\IN{{\hbox{{\rm I}\kern-.2em\hbox{\rm N}}}}
\def\IC{{\ \hbox{{\rm I}\kern-.6em\hbox{\bf C}}}}

\def\IZ{{\hbox{{\rm Z}\kern-.4em\hbox{\rm Z}}}}
\def\to{\rightarrow}
\def\d{{\rm d}}
\def\underarrow#1{\vbox{\ialign{##\crcr$\hfil\displaystyle
{#1}\hfil$\crcr\noalign{\kern1pt
\nointerlineskip}$\longrightarrow$\crcr}}}
%
\def\d{{\rm d}}
\def\ltorder{\mathrel{\raise.3ex\hbox{$<$}\mkern-14mu
             \lower0.6ex\hbox{$\sim$}}}
\def\lesssim{\mathrel{\raise.3ex\hbox{$<$}\mkern-14mu
             \lower0.6ex\hbox{$\sim$}}}
\def\Z{{\bf Z}}
\def\C{{\bf C}}

\def\N{{\bf N}}

\def\overlrarrow#1{\vbox{\ialign{##\crcr
$\leftrightarrow$\crcr\noalign{\kern-1pt\nointerlineskip}
$\hfil\displaystyle{#1}\hfil$\crcr}}}

\input phyzzx
\singlespace
\overfullrule=0pt

\def\bar{\overline}
\def\w{\widehat}
\tolerance=5000
\overfullrule=0pt
\twelvepoint

\pubnum{IASSNS-HEP-93/41}
\date{December, 1993}
\titlepage
\title{THE VERLINDE ALGEBRA AND THE COHOMOLOGY OF THE GRASSMANNIAN}
\vglue-.25in
\author{Edward Witten
\foot{Research supported in part by NSF Grant
PHY92-45317.}}
\medskip
\address{School of Natural Sciences
\break Institute for Advanced Study
\break Olden Lane
\break Princeton, NJ 08540}
\bigskip
\abstract{The article is devoted to a quantum field theory
explanation of the relationship
between the Verlinde algebra of the group $U(k)$ at level $N-k$
and the ``quantum''
cohomology of the Grassmannian of complex $k$ planes in $N$ space.
In \S2, I explain the relation between the Verlinde algebra and
the gauged WZW model of $G/G$; in \S3, I describe the quantum
cohomology and its origin in a quantum field theory; and in \S4,
I present a path integral argument for mapping between them.}
\endpage

\chapter{Introduction}

\def\d{d}
\def\D{{D}}
\REF\gepner{D. Gepner,  ``Fusion Rings And Geometry,'' Commun. Math. Phys.
{\bf 141} (1991) 381; D. Gepner and A. Schwimmer, ``Symplectic Fusion
Rings And Their Metric,'' Nucl. Phys. {\bf B380} (1992) 147; D. Gepner,
``Foundations Of Rational Conformal Field Theory, I'' (Cal Tech preprint,
1992).}
My main goal in these lecture notes will be to elucidate a formula
of Doron Gepner [\gepner],
which relates two mathematical objects, one rather old and one rather new.
Along the way we will consider a few other matters as well.

\def\N{{\cal N}}
\REF\vafa{C. Vafa, ``Topological Mirrors And Quantum Rings,''
in {\it Essays On Mirror Manifolds}, ed. S.-T. Yau (International
Press, 1992).}
\REF\intrilligator{K. Intriligator, ``Fusion Residues,'' Mod. Phys.
Lett. {\bf A6} (1991) 3543.}
The old structure is the cohomology ring of
the Grassmannian $G(k,N)$ of complex $k$ planes in $N$ space -- except
that one considers the quantum cohomology (or Floer instanton homology)
rather than
the classical cohomology.  The new
structure is the Verlinde algebra, which computes the Hilbert polynomial
of the moduli space of vector bundles on a curve.  Gepner's formula, as
we will consider it here,
\foot{Gepner actually discusses the classical cohomology of the Grassmannian
and identifies it with a close cousin of the Verlinde algebra.  The
refinement of Gepner's formula that we will consider was conjectured
by Vafa [\vafa] and Intriligator [\intrilligator].}
says that the quantum cohomology ring of $G(k,N)$ coincides
with the Verlinde algebra of the group $U(k)$ essentially
at level $N-k$.\foot{Actually if one decomposes the Lie algebra of
$U(k)$ as $su(k)\times u(1)$, then the level is $(N-k,N)$, that
is level $N-k$ for the $su(k)$ factor and level $N$ for the $u(1)$
factor.  The source of this subtlety will become clear in \S4.6.}

Gepner discovered his formula by computing the left and right hand side
and observing that they were equal.  We will seek a more conceptual
explanation, by representing the quantum cohomology ring of the Grassmannian
in a quantum field theory and reducing that quantum field theory
at low energies to another quantum field theory which is known to
compute the Verlinde algebra.

\REF\gerasimov{A. Gerasimov, ``Localization In GWZW And Verlinde Formula,''
hepth@xxx/9305090.}
\REF\blau{M. Blau and G. Thompson, ``Derivation Of The Verlinde Formula
{}From Chern-Simons Theory And The $G/G$ Model,'' Nucl. Phys. {\bf B408}
(1993) 345.}
The Verlinde formula appears in several quantum field theories.
Of these, the one that is relevant here
is the gauged WZW model, of $G/G$.  The shortest
and most complete explanation of its relation
to the Verlinde formula is due to Gerasimov [\gerasimov],
and I will explain his argument in \S2.
Part of the charm of the $G/G$ model is that it can be abelianized,
that is, reduced to a theory in which the gauge group is the maximal
torus of $G$, extended by the Weyl group [\blau].
The argument is simple in concept and will be summarized in \S2.6.

\REF\batyrev{V. Batyrev, ``Quantum Cohomology Rings Of Toric
Manifolds'' (preprint, 1993).}
In \S3, I explain at a qualitative level how the quantum cohomology
of the Grassmannian is represented in a quantum field theory,
and some general techniques for studying this field theory
and reducing it to a problem in gauge theory.
In \S4, I describe the arguments in more technical detail.
The analysis actually should be adaptable to other manifolds
that can be realized as symplectic quotients of linear spaces,
such as flag manifolds and toric varieties.  (The quantum cohomology
of a toric variety has been studied by Batyrev [\batyrev]; that of
a general flag manifold has apparently not yet been studied.)

\REF\div{A. D'Adda, M. Luscher, and P. DiVecchia, ``A $1/N$ Expandable
Series Of Nonlinear Sigma Models With Instantons,'' Nucl. Phys. {\bf B146}
(1978) 63, ``Topology And Higher Symmetries Of The Two Dimensional
Nonlinear Sigma Model,'' Phys. Report {\bf 49} (1979) 239.}
\REF\oldwit{E. Witten, ``Instantons, The Quark Model, And The $1/N$
Expansion,'' Nucl. Phys. {\bf B149} (1979) 285.}
\REF\brazil{E. Abdalla, M. Forger, and A. Lima Santos, ``Non-Local
Charges For Nonlinear Sigma Models On Grassmann Manifolds,'' Nucl. Phys.
{\bf B256} (1985) 145.}
\REF\cecotti{S. Cecotti and C. Vafa, ``On Classification Of $N=2$
Supersymmetric Theories,'' Harvard preprint HUTP-92-A064.}
This paper, despite its length, is based on an idea that can be
described very simply.
The two dimensional supersymmetric sigma
model with target $G(k,N)$ can be described as a $U(k)$ gauge
theory (in $\N=2$ superspace) with $N$ multiplets of chiral superfields
in the fundamental representation of $U(k)$.  It was studied from
this point of view in the case of $k=1$ (that is ${\bf CP}^{N-1}$) many years
ago [\div,\oldwit], and the generalization to arbitrary $k$ is also
familiar [\brazil,\cecotti].  At low energy, a suitable
$U(k)$ gauge theory with the matter content just stated
reduces to the supersymmetric sigma
model of the Grassmannian.  On the other hand, integrating out the $N$
matter multiplets, one gets an effective action for the $U(k)$ gauge
multiplet.  Because of a sort of mixing between scalars and vectors,
this low energy effective action has no massless particles; this is how
the presence of a mass gap has been shown in the past.  The novelty
in the present paper is simply the observation that the low energy effective
action is in fact a gauged WZW model of $U(k)/U(k)$.  Under this
low energy reduction, the topological correlation functions of the
sigma model -- which compute the quantum cohomology of $G(k,N)$ -- are
mapped into correlation functions of the $U(k)/U(k)$ model
that can be computed (as we recall in \S2) in terms of the Verlinde
algebra.  This gives the map between the two theories.

\REF\bertram{A. Bertram, G. Daskapoulos, and R. Wentworth,
``Gromov Invariants For Holomorphic Maps From Riemann Surfaces To
Grassmannians,'' preprint (April, 1993).}
\REF\lerche{W. Lerche, C. Vafa, and N. Warner, ``Chiral Rings In
$\N=2$ Superconformal Theories,'' Nucl. Phys. {\bf B324} (1989) 427.}
\REF\bourdeau{M. Bourdeau, E. J. Mlawer, H. Riggs, and H. J. Schnitzer,
``Topological Landau-Ginzburg Matter From $SP(N)_K$ Fusion Rings,''
Mod. Phys. Lett. {\bf A7} (1992) 689.}
The quantum cohomology of the Grassmannian has also been studied -- using,
more or less, a classical version of the same setup we will follow -- by
Bertram, Daskapoulos, and Wentworth [\bertram].
And there is a forthcoming mathematical
approach to Gepner's formula in work of Braam and Agnihorti.
The cohomology of the Grassmannian is closely related
to the chiral ring of a certain $\N=2$ superconformal field theory
[\lerche] (somewhat misleadingly called a $U(N)/U(k)\times U(N-k)$
coset model); this model probably should be included in the story,
but that will not be done here.  Some of the phenomena we will study
have analogs for real and symplectic Grassmannians, as in [\bourdeau]
and the second paper cited in [\gepner]; it would be interesting
to try to extend the analysis for those cases.

\S2 and \S3 can be read independently of one another.  \S4 requires
more familiarity with methods of physics than either \S2 or \S3.
Physicists may want to start with \S4.

\chapter{The Verlinde Formula And The $G/G$ Model}

First of all, the Verlinde algebra counts theta functions, such
as the classical theta functions of Jacobi and their generalizations.
In modern language, the classical theta functions can be described as
follows.  Let ${\cal T}$ be a complex torus of dimensions $g$, ${\cal L}$
a line bundle defining a principal polarization, and $s$ a positive
integer.  Then the space of level $s$ theta functions is $H^0({\cal T},
{\cal L}^{\otimes s})$.  The dimension of this space can be readily determined
from the Riemann-Roch theorem.

For example, ${\cal T}$ might be the Jacobian ${\cal J}$ of a complex
Riemann surface $\Sigma$, that is, the moduli space of holomorphic
line bundles over $\Sigma$ of some given degree.  This example suggests
the generalization to the ``non-abelian theta functions'' of A. Weil.
Here one replaces the Jacobian of $\Sigma$ by the moduli space ${\cal R}$
of rank $k$ (stable) holomorphic vector bundles over $\Sigma$; now
a ``non-abelian theta function'' at level $s$ is an element
of $H^0({\cal R},{\cal L}^{\otimes s})$.  Though the Riemann-Roch
theorem gives a formula for the dimension of this space, this formula
is difficult to use in the non-abelian case
as it involves invariants of ${\cal R}$
that are not easy to determine directly.

\REF\bott{R. Bott, ``On E. Verlinde's Formula In The Context Of
Stable Bundles,'' in {\it Topological Methods In Quantum Field
Theories}, ed. W. Nahm et. al. (World Scientific, 1991).}
The Verlinde algebra gives on the other hand a practical formula
for the dimension of $H^0({\cal R},{\cal L}^{\otimes s})$; the formula
was described very explicitly by Raoul Bott in [\bott].  Roughly speaking,
the origin of the Verlinde formula in differential geometry is as follows.
Via Hodge theory, ${\cal R}$ is endowed with a natural Kahler metric.
As the dimension of the space of non-abelian theta functions is
independent of the complex structure of $\Sigma$, one
can choose the complex structure to simplify the problem.
It is convenient to take $\Sigma$ to be
a nearly degenerate surface consisting of three-holed
spheres joined by long tubes.   Using the behavior of the differential
geometry of ${\cal R}$ in this limit, one can write $H^0({\cal R},{\cal L}
^{\otimes s})$ as a sum of tensor products of
similar spaces for a three-holed sphere
with some branching around the holes.  The Verlinde algebra encodes
the details of this.

The Verlinde algebra arises in several quantum field theories:

\REF\everlinde{E. Verlinde, ``Fusion Rules And Modular Transformations
In 2-D Conformal Field Theory,'' Nucl. Phys. {\bf B300} (360) 1988.}
(1) It originally arose
[\everlinde] in the WZW model, a conformal field theory
(whose Lagrangian we will recall later) that
governs maps from a Riemann surface
$\Sigma$ to a compact Lie group $G$.
Mathematically, this model is related to representations of affine
Lie algebras, the unitary action of the modular group $SL(2,{\bf Z})$ on their
characters, etc.

(2)  The Verlinde formula is an important ingredient in understanding
Chern-Simons gauge theory on a three-manifold.  Thus it is relevant
to the knot and three-manifold invariants constructed from quantum
field theory.

(3)  The Verlinde formula also enters in the gauged WZW model,
governing a pair $(g,A)$, where $A$ is a connection on a principal
$G$ bundle $P$ over a Riemann surface, and $g$ is a section of $P\times_G G$
(where $G$ acts on itself via the adjoint action).

\REF\gawedzki{K. Gawedzki and A. Kupianen, ``A $G/H$ Conformal Field
Theory From Gauged WZW Models,'' Phys. Lett. {\bf 215B} (1988) 119,
``Coset Construction From Functional Integrals,'' Nucl. Phys.
{\bf B320(FS)} (1989) 649; K. Gawedzki, ``Constructive Conformal Field
Theory,'' in {\it Functional Integration, Geometry, And Strings},
eds. Z. Hava and J. Sobczyk (Birkhauser, 1989).}
\REF\uggwitten{E. Witten, ``On Holomorphic Factorization Of WZW And
Coset Models,'' Commun. Math. Phys. {\bf 114} (1992) 189.}
Of these it is
the third -- which was discovered most recently -- that will enter
our story.  The present
section is therefore mainly devoted to an explanation -- following
Gerasimov [\gerasimov], who reinterpreted earlier formulas
[\gawedzki,\uggwitten] -- of the gauged WZW model and its relation
to nonabelian theta functions.

\section{Gauge Theory And The Prequantum Line Bundle In Two Dimensions}

\REF\axelrod{S. Axelrod, ``Geometric Quantization Of Chern-Simons
Gauge Theory,'' Ph.D. Thesis, Princeton University (1991).}
\REF\gaw{K. Gawedzki, ``Topological Actions In Two-Dimensional
Quantum Field Theories,'' in {\it Non-perturbative Quantum Field
Theory,} ed. G. 't Hooft (Plenum Press, 1988); G. Felder, K.
Gawedzki, and A. Kupianen, ``Spectra of Wess-Zumino-Witten Models
With Arbitrary Simple Groups,'' Commun. Math. Phys. {\bf 117} (1988) 127.}
\REF\abott{M. F. Atiyah and R. Bott, ``The Yang-Mills Equations Over
Riemann Surfaces,'' Philos. Trans. R. Soc. London. {\bf A308} (1982) 523.}
\def\A{{\cal A}}
Let $G$ be a compact Lie group, $\Sigma$ a closed oriented two-manifold without
boundary,
and $P$ a principal $G$ bundle over $\Sigma$.
To achieve some minor simplifications in the exposition, I will suppose
$G$ simple, connected, and simply connected.  (Notation aside, the only
novelty required to treat a general compact Lie group is that more
care is required in defining the functional $\Gamma(g,A)$ that
appears below; see [\axelrod, \S4].)
One consequence of the assumption about $G$ is that $P$ is trivial.

Let $\A$ be the space
of connections on $P$.  $\A$ has a natural symplectic structure $\omega$ that
can be defined with no choice of metric or complex structure on $\Sigma$.
(This and some other facts that I summarize presently are originally
due to Atiyah and Bott [\abott].)  The symplectic structure
can be defined by the formula
$$\omega(a_1,a_2)={1\over 2\pi}\int_\Sigma\Tr a_1\wedge a_2,\eqn\defform$$
where $a_1$ and $a_2$ are adjoint-valued one-forms representing
tangent vectors to $\A$.  Here $\Tr$ is an invariant quadratic form
on the Lie algebra of $G$, defined for $G=SU(k)$ to be the trace in the
$k$ dimensional representation; in general one can take $\Tr$ to
be the smallest positive multiple of the trace in the adjoint
representation such that the differential form $\Theta$ introduced
below has periods that are multiples of $2\pi$.

A prequantum line bundle ${\cal L}$ over
${\cal A}$ is a unitary line bundle with
a connection of curvature $-i\omega$.  ${\cal L}$ exists and is unique
up to isomorphism since ${\cal A}$ is an affine space.
We can take ${\cal L}$ to be the trivial bundle
with a connection defined by the following formula:
$${D\over DA_i}={\delta\over\delta A_i}+{i\over
4\pi}\epsilon^{ij}A_j.\eqn\bbu$$
($\epsilon^{ij}$ is the Levi-Civita antisymmetric tensor; when
local complex coordinates are introduced, we will take $\epsilon^{z\bar z}
=-\epsilon^{\bar z z}=i$.)
The $k^{th}$ power ${\cal L}^{\otimes k}$ is therefore the trivial
bundle endowed with the connection
$${D\over DA_i}={\delta\over\delta A_i}+{ik\over 4\pi}\epsilon^{ij}A_j.
\eqn\ccub$$

Let $\w G$ be the group of gauge transformations.  If $P$ is trivialized,
a gauge transformation is a map $g:\Sigma\to G$,
and acts on the connection $A$ by
$$ A \to A^g=gAg^{-1}-\d g \cdot g^{-1}. \eqn\mcon$$
More invariantly, $g$ is a section of $P\times_G G$, where $G$ acts on
itself in the adjoint representation,  and the action of
$g$ on ${\cal A}$ should be written as
$$\d_A\to
g \d_A g^{-1}, \eqn\eeqn$$
with $\d_A$ the gauge-covariant extension of the exterior
derivative.  At the Lie algebra level this is
$$ A\to A-\d_A\alpha, \eqn\con$$
where $\alpha$ is a section of $P\times_G\bf g$, with $\bf g$ being
the Lie algebra of $G$, on which $G$ acts by the adjoint action.

The action of the gauge group on the space ${\cal A}$ of connections
lifts to an action on the prequantum line bundle.  At the Lie algebra
level, the lift is generated by the operators
$$ D_i{D\over DA_i}-{ik\over 4\pi}\epsilon^{ij}F_{ij},\eqn\hocco$$
with $F=\d A+A\wedge A$ the curvature form.  \hocco\ means very concretely
that the infinitesimal gauge transformation \con\ is represented
on sections of ${\cal L}$ by the operator
$$\int_\Sigma \Tr\alpha\left(D_i{D\over DA_i}-{ik\over 4\pi}F\right).
            \eqn\bocco$$

\REF\weitsinger{T. R. Ramadas, I. M. Singer, and J. Weitsman, ``Some
Comments On Chern-Simons Gauge Theory,'' Commun. Math. Phys. {\bf 126} (1989)
409.}
\REF\wwwitten{E. Witten, ``Global Aspects Of Current Algebra,'' Nucl.
Phys. {\bf B223} (1983) 422, ``Non-Abelian Bosonization In Two
Dimensions,'' Commun. Math. Phys. {\bf 92} (1984) 455.}
\REF\wz{J. Wess and B. Zumino, ``Consequences Of Anomalous Ward Identities,''
Phys. Lett. {\bf 37B} (1971) 95.}
\REF\polyakov{A. M. Polyakov and P. B. Wiegmann, ``Theory Of Non-Abelian
Goldstone Bosons In Two Dimensions,'' Phys. Lett. {\bf B131} (1983) 121.}
Even globally, at the group level, the $\w G$ action on ${\cal L}$
can be described rather explicitly [\weitsinger].
Pick a three
manifold $B$ with $\partial B=\Sigma$.  Given $g:\Sigma\to G$,
extend $g$ to a map (which I will also call $g$) from $B$ to $G$.
(The extension exists because of our assumption that $\pi_0(G)=\pi_1(G)=0$.
See [\gaw,\axelrod] for the definition of $\Gamma$ without
this simplifying assumption.)
Define as in [\wwwitten]
$$ \Gamma(g)={1\over 12 \pi}\int_B\Tr g^{-1}\d g\wedge g^{-1}\d g\wedge
g^{-1}\d g,  \eqn\wzfun$$
which is known as the
Wess-Zumino anomaly functional [\wz].
This is equivalent to
$$\Gamma(g)=\int_Bg^*(\Theta),\eqn\pooo$$
where
$$\Theta={1\over 12\pi}\Tr g^{-1}\d g\wedge g^{-1}\d g\wedge g^{-1}\d g
\eqn\sonnop$$
is a left- and right-invariant closed three-form on $G$.
The periods of $\Theta$ are multiples of $2\pi$.  This
ensures that, regarded as
a map to ${\bf R}/2\pi {\bf Z}$, $\Gamma(g)$ depends only on the restriction
of $g$ to $\Sigma$.  Note that $\Gamma$ is defined purely in differential
topology; no metric or complex structure on $\Sigma$ is required.

It follows rather directly from the definition of
$\Gamma$ that for $g,h:\Sigma\to G$,
$$\Gamma(gh)=\Gamma(g)+\Gamma(h)-{1\over 4\pi}\int_\Sigma\Tr g^{-1}\d g
\wedge \d h\cdot h^{-1}. \eqn\pwform$$
A variant of this equation is called the Polyakov-Wiegmann formula
[\polyakov].

Now, given a connection $A$ on $P$, set
$$W(g,A)=\Gamma(g)-{1\over 4\pi}\int_\Sigma\Tr A\wedge g^{-1}\d g.\eqn\mcin$$
{}From \pwform, it follows almost immediately that
$$W(gh, A)= W(g, A^h)+W(h,A). \eqn\cform$$
{}From this we can define an action of the gauge group $\widehat G$ on
the space of functions of $A$.  In fact, setting
$$g^*\chi(A)= \exp(ikW(g,A))\cdot \chi(A^g), \eqn\jform$$
we have $(gh)^*=h^*g^*$.  Differentiating \jform\ with respect to $g$
at $g=1$, one sees that this particular lift induces \hocco\ at
the Lie algebra level; so \jform\ is the desired lifting of the action
of the gauge group to an action on the prequantum line bundle ${\cal L}$.

\section{Non-Abelian Theta Functions}

So far we have considered $\Sigma$ simply as a closed, oriented two-manifold
without boundary.  If one picks a complex structure on $\Sigma$,
some additional interesting constructions can be made [\abott].
A complex structure
on $\Sigma$ induces a complex structure on the space ${\cal A}$ of connections.
One simply declares that the $(0,1)$ part of $A$ is holomorphic
and the $(1,0)$ part is antiholomorphic.
If $z,\bar z$ are local complex coordinates on $\Sigma$,
then the connection \ccub\ characterizing ${\cal L}^{\otimes k}$ can be written
$$\eqalign{{D\over DA_z} & ={\delta\over\delta A_z}-{k\over 4\pi}A_{\bar z}
 \cr
{D\over DA_{\bar z}} & ={\delta\over\delta A_{\bar z}}+
{k\over 4\pi}A_{z}.
 \cr} \eqn\roar$$
The complex structure on ${\cal A}$
can be described in very down-to-earth terms by saying that
a holomorphic function on ${\cal A}$ is a function annihilated
by $\delta/\delta A_z$.  Correspondingly,
a holomorphic section of ${\cal L}^{\otimes k}$
is a section annihilated by $D/D A_z$.  Even more explicitly, a holomorphic
section of ${\cal L}^{\otimes k}$
is a function $\chi(A_z,A_{\bar z})$ which can be written
$$\chi(A_z,A_{\bar z})=\exp\left({k\over 4\pi}\int_\Sigma\Tr A_zA_{\bar z}
\right)\cdot \w \chi(A_{\bar z}), \eqn\oncoo$$
with $\w \chi(A_{\bar z})$ an ordinary holomorphic function on ${\cal A}$.

Once a complex structure is picked on $\Sigma$,
the connection $A$ determines operators $\bar\partial_A$ giving
complex structures to vector bundles $P\times_G\bf r$, with ${\bf r}$
a representation of $G$.  The action of gauge
transformations on $A$ can be described by the action on the
$\bar\partial_A$ operators:
$$\bar\partial_A\to g\cdot \bar\partial_A\cdot g^{-1}. \eqn\mnon$$
Since this formula makes sense for complex $g$, the $\w G$ action
on ${\cal A}$ extends to an action of the complexified gauge group
$\w G_{{\bf C}}$ (consisting of maps of $\Sigma$ to the complexification
$G_{\bf C}$ of $G$).

Two $\bar\partial_A$ operators define equivalent holomorphic bundles
if and only if they are related as in \mnon.  So the quotient
${\cal A}/G_{\bf C}$ (in case the $G_{\bf C}$ action
is not free, the quotient must be taken in the sense of geometric invariant
theory) is the same as the moduli space ${\cal R}$ of (stable)
holomorphic principal $G$ bundles over $\Sigma$.

The formulas used to describe the lift of the $\w G$ action to ${\cal L}$
make sense when $g$ is complex, so we get a lift of the
$\w G_{\bf C}$ action to ${\cal L}$.  One defines a line
bundle over ${\cal R}$ -- which we will also call ${\cal L}$ --
whose sections over an open set $U\subset {\cal R}$
are the same as the $\w G_{\bf C}$-invariant sections
of ${\cal L}$ over the inverse image of $U$ in ${\cal A}$.

\subsection{Non-Abelian Theta Functions}

The space of non-abelian theta functions, at level $k$, is
$H^0({\cal R},{\cal L}^{\otimes k})$.  From what has just been said, this
is the same as the $\w G$-invariant (or equivalently, $\w G_{\bf C}$-invariant)
subspace ${\cal H}^{\w G}$ of ${\cal H}=H^0({\cal A},{\cal L}^{\otimes k})$.
We want to determine the dimension of ${\cal H}^{\w G}$.

The strategy, as in [\gerasimov],
will be as follows.  We will find a very convenient description
of the action of $\w G$ on ${\cal H}$.  In fact, for $g\in \w G$,
we will find an explicit integral kernel $K(A,B;g)$ (with $A,B\in {\cal A}$)
such that for $\chi\in {\cal H}$,
$$g^*\chi(A)= \int DB  \,\,\, K(A,B;g)\chi(B).  \eqn\mokno$$
(For fixed $g$, $K(A,B;g)$ is a section of $p_1^*({\cal L}^{\otimes k})\otimes
p_2^*({\cal L}^{\otimes(-k)})$ over ${\cal A}\times {\cal A}$,
with $p_1$ and $p_2$ being the two projections to ${\cal A}$.)
Here $DB$ is the natural symplectic measure, normalized
in a way that will be specified in \S2.4,
on the symplectic manifold ${\cal A}$.

Now the projection operator $\Pi:{\cal H}\to {\cal H}^{\widehat G}$
can be written
$$\Pi={1\over {\rm vol}(\w G)}\int_{\w G}Dg \,\,\,g^*.\eqn\jory$$
Here formally  $Dg$ is a Haar measure on $\w G$ and ${\rm vol}(\w G)$ is the
volume of $\w G$ computed with the same measure.
The dimension of ${\cal H}^G=H^0({\cal R},{\cal L}^{\otimes k})$
is the same as $\Tr \Pi$, and so can
evidently be written
$${\rm dim}\, H^0({\cal R},{\cal L}^{\otimes k})
={1\over {\rm vol}(\w G)}\int Dg \,\,DA \,\,\,\,\,  K(A,A;g).
\eqn\evwri$$
It remains to construct a suitable kernel $K$.
This will be done using gauged WZW models.

\section{Gauged WZW Models}

Let $\Sigma$ be a complex Riemann surface; the complex structure
determines the Hodge duality operator $*$ on one-forms.
For a map $g:\Sigma\to G$, the WZW functional
is
$$I(g)=-{1\over 8\pi}\int_\Sigma \Tr g^{-1}\d g\wedge * g^{-1}\d g
-i\Gamma(g), \eqn\qzwlag$$
where $\Gamma(g)$ was defined in \wzfun.  While $\Gamma(g)$ is defined
purely in differential topology, the first term in the definition of $I(g)$
depends on the complex structure of $\Sigma$ through the $*$ operator.
The quantum field theory
with Lagrangian $L(g)=kI(g)$
is conformally invariant and describes level $k$ highest weight
representations of the loop group of $G$ and the action of the modular
group on their characters.

$G\times G$ acts on $G$ by left and
right multiplication ($g\to a g b^{-1}$).  Let us denote this copy of
$G\times G$  as $G_L\times G_R$ with $G_L$ and $G_R$ acting on the left
and right respectively.
$I(g)$ is invariant under $G_L\times G_R$.  We want to pick a subgroup
$H\subset G_L\times G_R$ and construct a gauge invariant extension
of $I(g)$ with gauge group $H$.  What this means is that we introduce
a principal $H$ bundle $P$, with connection $A$, and we replace the map
$g:\Sigma\to G$ by a section of the bundle $P\times_HG$; here $G$ is
understood as the trivial principal $G$ bundle over $\Sigma$, and $H$ acts
on $G$ via its chosen embedding in $G_L\times G_R$.  We want
to construct a natural, gauge invariant functional $I(g,A)$ that
reduces at $A=0$ to $I(g)$.

There is no problem in constructing a gauge invariant extension of the first
term in \qzwlag.  One simply replaces the exterior derivative by its
gauge-covariant extension:
$$-{1\over 8\pi}\int_\Sigma\Tr g^{-1}\d_A g\wedge * g^{-1}\d_Ag. \eqn\jopipo$$
On the other hand, there is a topological obstruction to constructing
a gauge invariant extension of $\Gamma(g)$.  The requirement is that
the class in $H^3(G,{\bf Z})$ that determines the functional $\Gamma$
(and which in real cohomology is represented by the differential
form $\Theta$) should have an extension to the equivariant cohomology
group $H^3_H(G,{\bf Z})$.  This is explained in [\axelrod,\S4];
a  quick explanation at the level of de Rham theory
(ignoring the torsion in $H^3_H(G,{\bf Z})$)
is in the appendix of [\uggwitten].

\REF\gko{P. Goddard, A. Kent, and D. Olive, ``Virasoro Algebras And
Coset Space Models,'' Phys. Lett. {\bf 152B} (1985) 88.}
\REF\rabinovici{A. Altschuler, K. Bardacki, and E. Rabinovici,
``String Models With $c<1$ Components,'' Nucl. Phys. {\bf B299} (1988) 157,
A. Altschuler, K. Bardacki, and E. Rabinovici,
``A Construction Of the $c<1$ Modular Invariant Partition Functions,''
Commun. Math. Phys. {\bf 118} (1988) 241.}
\REF\schnitzer{D. Karabali, and H. J. Schnitzer, ``BRST Quantization Of
The Gauged WZW Action And Coset Conformal Field Theories,'' Nucl. Phys.
{\bf B329} (1990) 649; D. Karabali, Q-Han Park, H. J. Schnitzer, and
Zhu Yang,
``A GKO Construction Based On A Path Integral Formulation Of Gauged
Wess-Zumino-Witten Actions,'' Phys. Lett. {\bf B216} (1989) 307.}
As explained, for instance, in that appendix, the condition for existence
of a gauge invariant extension of $\Gamma(g)$
can be put in the following very explicit form.  If $T_a,\,\,\,a=1\dots
\dim(H)$ are a basis of the Lie algebra of $H$, and if the embedding
$H\subset G_L\times G_R$ is described at the Lie algebra level by
$T_a\to (T_{a,L},T_{a,R})$, then the requirement is
$$\Tr T_{a,L}T_{b,L}=\Tr T_{a,R}T_{b,R}, ~~~ {\rm for~all}~a,b.\eqn\forall$$
A subgroup $H\subset G_L\times G_R$ obeying this condition is said to
be anomaly-free.
For such an $H$, the gauge invariant extension of $\Gamma$ exists and is
explicitly
$$\eqalign{\Gamma(g,A)=&
\Gamma(g)-{1\over 4\pi}\sum_a\int_\Sigma A^a\Tr\left(T_{a,L} \d g\cdot
g^{-1}+T_{a,R}g^{-1}\d g\right)\cr &-{1\over 8\pi}\sum_{a,b}
\int A^a\wedge A^b\Tr
\left(T_{a,R}g^{-1}T_{b,L}g-T_{b,R}g^{-1}T_{a,L}g\right)
.\cr}\eqn\qqq$$
Combining these formulas, one gets for anomaly-free $H$ a gauge invariant
extension of the WZW functional,
$$I(g,A)=
-{1\over 8\pi}\int_\Sigma\Tr g^{-1}\d_A g\wedge * g^{-1}\d_Ag-i\Gamma(g,A).
\eqn\jopipol$$
The quantum field theories with Lagrangians $L(g,A)=kI(g,A)$, $k$ a positive
integer, are called $G/H$ models.
\foot{The terminology is somewhat misleading since these models are not
the most obvious sigma models with target space $G/H$; and one is not
allowed to use the most obvious $H$ actions on $G$, such as the left
or right actions, which are anomalous.  The terminology is used because
the models are believed [\rabinovici,\schnitzer,
\gawedzki,\uggwitten] to be equivalent
to GKO models [\gko], which were originally described algebraically,
and are conventionally called $G/H$ models or coset models.
The claimed equivalence to the GKO models implies in particular that
the models are conformally invariant at the quantum level.}
Note that for given $G$ and $H$, there may be several $G/H$ models,
since there may be several anomaly-free embeddings of $H$ in $G_L\times G_R$.

If $H$ is any subgroup of $G$, then the diagonal embedding of $H$ in
$G_L\times G_R$ is always anomaly free.  The model determined by
such a diagonal embedding is often called ``the'' $G/H$ model.
If we pick local complex coordinates $z,\bar z$ on $\Sigma$ (which
will facilitate a small calculation needed presently) and write
the measure $|\d z\wedge \d\bar z|$ as $\d^2z$, then the Lagrangian of
the diagonal $G/H$ model is explicitly $k$ times
$$\eqalign{
I(g,A)=&
I(g)-{1\over 2\pi}\int_\Sigma\d^2 z\Tr A_z\partial_{\bar z}
g\cdot g^{-1}\cr &
+{1\over 2\pi}\int_\Sigma\d^2z\Tr A_{\bar z}g^{-1}\partial_zg
-{1\over 2\pi}\int_\Sigma\d^2z\Tr\left(A_z
A_{\bar z}-A_zgA_{\bar z}g^{-1}\right)
        .\cr }\eqn\juryfile$$

Our interest will center on the special case of the diagonal $G/H$ model
for $H=G$.  This is then the $G/G$ model with adjoint action of $G$ on
itself.  The Lagrangian is $k$ times \juryfile, and the partition function
at level $k$ is
$$Z_k(G,\Sigma)=
 {1\over {\rm vol}(\w G)}\int \D g\,\,\,\D A \,\,\,\,\,\exp\left(
-kI(g,A)\right).   \eqn\niso$$
Our goal is to use \evwri\ to show that
$Z_k(G,\Sigma)$ coincides with the dimension of the space of non-abelian theta
functions at level $k$.

\section{The Kernel}

One more special case is important: $H=G_L\times G_R$.
This is an anomalous subgroup, so there is no gauge invariant
$G/H$ Lagrangian and no $G/H$ quantum field theory for this $H$.
We will do something else instead.

Denote the $G_L$ and $G_R$ components of an $H$ connection as $A$ and $B$. Set
$$I(g,A,B)=I(g)+{1\over 2\pi}\int \d^2z\Tr\left( A_{\bar z}g^{-1}\partial_zg
-B_z\partial_{\bar z}g\cdot g^{-1}+B_zgA_{\bar z} g^{-1}
-{1\over 2}A_zA_{\bar z}-{1\over 2}B_zB_{\bar z}\right). \eqn\ipo$$
This functional is determined by the following: it
is not gauge invariant, but its change under
a gauge transformation is independent of $ g$ and related in a useful
way to the geometry of the prequantum line bundle.
In fact, under an infinitesimal
gauge transformation
$$\delta g=vg-gu,~~~\delta A=-\d_Au,~~~\delta B=-\d_Av ,\eqn\infg$$
we have
$$\delta I(g,A,B)={i\over 4\pi}\int_\Sigma\Tr\left(u\,\d A-v\,\d B\right).
\eqn\ginf$$
(The fact that an extension  $I(g,A,B)$ of $I(G)$ exists
with these properties has a conceptual explanation noted
in the appendix to [\uggwitten].)

Now, set
$$K(A,B;g)=\exp\left(-kI(g,A,B)\right).   \eqn\cnxon$$
In its dependence on $A$, $K$ can be interpreted as a holomorphic
section of ${\cal L}^{\otimes k}$; this just means that $K$ is
independent of $A_z$ except for the exponential factor prescribed in
\oncoo.  Likewise, in its dependence on $B$, $K$ is an anti-holomorphic
section of ${\cal L}^{\otimes(-k)}$; this means that it is independent
of $B_{\bar z}$ except for a similar exponential. In [\uggwitten], the above
facts were used to describe holomorphic factorization
of WZW and coset models.  Gerasimov's insight [\gerasimov]
was that $K$ is actually the kernel representing the action of the
gauge group on ${\cal H}=H^0({\cal A},{\cal L}^{\otimes k})$.
This means that for $\chi\in {\cal H}$ and $g\in \w G$,
$$g^*\chi(A)=\int \D B\,\,\,\,\,K(A,B;g) \chi(B).  \eqn\niuggo$$

To show this, we first as in \oncoo\ write the holomorphic section $\chi$
as
$$\chi(B)=\exp\left({k\over 4\pi}\int_\Sigma\d^2z\Tr B_zB_{\bar z}\right)
\w\chi(B_{\bar z}) \eqn\noko$$
with $\w\chi$ an ordinary holomorphic function.
The $B$-dependent factors in the integral
on the right hand side of \niuggo\ are
$$\int \D B\,\,\,\,\exp\left({k\over 2\pi}\int_\Sigma\d^2z
\Tr\left( B_zB_{\bar z}
-B_zA_{\bar z}{}^g\right)\right)\cdot \w \chi(B_{\bar z}). \eqn\goko$$
To perform such an integral, the basic fact is that if $f(\phi)$ is
a holomorphic function that grows at infinity more slowly than
$\exp(|\phi|^2)$, then
$${1\over \pi}\int_{\bf C} |d\phi\wedge d\bar\phi|
      \exp(-\bar\phi \phi +a\bar\phi)f(\phi)=f(a). \eqn\ompo$$
Using this fact and normalizing the symplectic
measure on ${\cal A}$ so that
$$\int \D B\,\,\exp\left({k\over 2\pi}\int_\Sigma
\d^2z\Tr B_zB_{\bar z}\right) = 1 \eqn\mpo$$
(to avoid a determinant that would otherwise arise in using \ompo), we get
simply
$$\int \D B\,\,\,\,\exp\left({k\over 2\pi}\int_\Sigma\d^2z
\Tr\left( B_zB_{\bar z}
-B_zA_{\bar z}{}^g\right)\right)\cdot \w \chi(B_{\bar z})=
\w\chi(A_{\bar z}^g). \eqn\gokko$$
The integral in \niuggo\ thereby becomes
$$\eqalign{\int \D B \,\,\,\,K(A,B;g)\chi(B)&
=\exp\left(-k\left(I(g)+{1\over 2\pi}
\int \d^2z\Tr A_{\bar z}g^{-1}\d g -{1\over 4\pi}
\int \d^2z\Tr A_zA_{\bar z}
\right)\right)\cr &~~~\cdot \w\chi(A_{\bar z}{}^g).\cr} \eqn\kson$$
Using \noko\ to reexpress $\w\chi$ in terms of $\chi$, and using
the explicit forms of $I(g)$ and $A^g$, we get
$$\int \D B\,\,\,K(A,B;g)\chi(B)=\exp\left(ik\left(\Gamma(g)-{1\over 4\pi}
\int_\Sigma\Tr A\wedge g^{-1}\d g\right)\right)\cdot \chi(A^g). \eqn\dson$$
Using the definition of $g^*$ in \jform, this indeed coincides
with the desired formula \niuggo.

As we saw in arriving at \evwri, it follows that the dimension of the
space of non-abelian theta functions is
$$\dim H^0({\cal R},{\cal L}^{\otimes k})={1\over {\rm vol}(\w G)}
\int \D g \,\,\D A\,\,\, K(A,A;g).
\eqn\snosno$$
But the integral on the right is precisely the partition function
\niso\ of the $G/G$ model (since $K(A,A;g)=\exp(-kI(g,A))$).
So we have arrived at the main goal of this section: identifying
the dimension of the space of non-abelian theta functions with the
partition function of the $G/G$ model.

\subsection{Inclusion Of Marked Points}

Now we would
like to extend the analysis slightly to the case of a Riemann surface
$\Sigma$ with marked points labeled by representations of $G$.
The $G/G$ model in this situation will give a path integral representation
of the Verlinde algebra.  (This generalization might be omitted on a first
reading.)

Suppose one has a represention $\rho$ of a compact Lie group $G$
in a Hilbert space ${\cal H}$.  Then as in \jory, the projection
operator onto the invariant subspace of ${\cal H}$ is
$$\Pi= {1\over {\rm vol}(G)}\int_G Dg \,\,\rho(g), \eqn\uxx$$
with $Dg$ an invariant measure on $G$ and ${\rm vol}(G)$ the
volume of $G$ computed with that measure.
The trace of $\Pi$ is the multiplicity with which the trivial
representation of $G$ appears in ${\cal H}$.

Now pick an irreducible representation
$V$ of $G$, that is a vector space $V$ in which $G$ acts
irreducibly by $g\to \rho_V(g)\in {\rm Aut}(V)$.  We want a formula
for the multiplicity with which $V$ appears in $G$.
We can reduce to the previous case as follows.  Let $\bar V$
be the dual or complex conjugate representation of $G$.  The multiplicity
with which $V$ appears in ${\cal H}$ is the same as the multiplicity
with which the trivial representation appears in ${\cal H}\otimes {\bar V}$.
So we define the projection operator $\Pi_{V}$ onto the $G$-invariant
subspace of ${\cal H}\otimes {\bar V}$:
$$\Pi_V={1\over {\rm vol}(G)}\int Dg\,\,\, \rho(g)\otimes \rho_{\bar V}(g).
\eqn\hudxo$$
The multiplicity with which $V$ appears in ${\cal H}$ is
$${\rm mult}(V)=\Tr \Pi_V.       \eqn\udxo$$

We want to apply this to the case in which $G$ is replaced by the
group $\widehat G$ of gauge transformations of a principal $G$ bundle
$P\to \Sigma$; and ${\cal H}$ will be, as above,
$H^0({\cal A},{\cal L}^{\otimes k})$.  The representations we will
use will be the following simple ones.
For a point $x\in \Sigma$, let $r_x:\widehat G\to G$ be the
map of evaluation at $x$.  For any representation $\rho_V:G\to {\rm Aut}(V)$
of $ G$, we have the corresponding representation $\rho_{x,V}
=\rho_V\circ r_x$ of $\widehat G$.
Pick now points $x_i\in \Sigma$, labeled by representations $V_i$,
and let $V=\otimes_iV_i$ with $\widehat G$ acting by
$$\rho_V=\otimes_i \rho_{x_i,V_i}.  \eqn\ingo$$
The conjugate representation is $\rho_{\bar V}=\otimes_i\rho_{x_i,\bar V_i}$.

We want to find a path integral representation of the multiplicity
with which $V$ appears in ${\cal H}$, along the lines of
\udxo.  To this aim we must calculate
$$\Tr\left(\rho(g)\otimes
\rho_{\bar V}(g)\right)=\Tr\rho(g)\cdot\Tr \rho_{\bar V}(g).\eqn\faf$$
Here the first factor has a path integral expression; in fact,
$$\Tr\rho(g)=\int DA \,\, K(A,A;g), \eqn\mcco$$
with $K(A,B;g)$ the kernel introduced in \cnxon.
The second factor is simply
$$\Tr\rho_{\bar V}(g) = \prod_i \Tr_{\bar V_i}g(x_i). \eqn\imoxox$$
So we get
$${\rm mult}(V)=\dim \left({\cal H}\otimes \bar V\right)^{\widehat G}
={1\over {\rm vol}(\widehat G)}\int Dg\,\,DA\,\,\exp(-kI(g,A))
\cdot\prod_i\Tr_{\bar V_i} g(x_i). \eqn\micnic$$
The right hand side is usually called the (unnormalized)
correlation function,
$$\left\langle \prod_i \Tr_{\bar V_i}g(x_i)\right\rangle\eqn\oxxo$$
in the gauged WZW model.  \oxxo\ would be unchanged if all $\bar V_i$
are replaced by $V_i$; the gauged WZW action has a symmetry (coming from an
involution of $G$ that exchanges all representations with their
complex conjugates) that ensures this.

\subsection{Relation To The Verlinde Algebra}

Now let us relate this to the Verlinde algebra.
Let $T$ be the maximal torus of $G$ and $G/T$ the quotient
of $G$ by the right action of $T$.  For any irreducible representation $V$
of $G$, there is a homogeneous line bundle ${\cal S}$ over
$G/T$ such that $ H^0(G/T,{\cal S})$ is isomorphic to $V$.

Given marked points $x_1,\dots, x_s$ on $\Sigma$, let
$\widehat {\cal A}$ be the symplectic manifold
$$\widehat{ \cal A}={\cal A}\times\prod_{i=1}^s(G/T)_i \eqn\oopp$$
where $(G/T)_i$ is a copy of $G/T$ ``sitting'' at $x_i$.
This is an informal way to say that the gauge group $\widehat G$
(and its complexification $\widehat G_{\bf C}$) acts on
$(G/T)_i$ by composition of the evaluation map $r_{x_i}$
with the natural action of $G$ (or $G_{\bf C}$) on $G/T$.

If we are given irreducible
representations $V_i$ of $G$, let for each $i$ ${\cal S}_i$
be a homogeneous line bundle over $(G/T)_i$ such that
$H^0((G/T)_i,{\cal S}_i)\cong \overline V_i$.
Define a homogeneous line bundle $\w {\cal L}$ over $\w{ \cal A}$
by
$$\w {\cal L}={\cal L}^{\otimes k}\otimes\left(\otimes_i{\cal S}_i\right).
\eqn\hins$$
(In an obvious way, I have identified the line bundles ${\cal L}$
and ${\cal S}_i$ with their pullbacks to $\w{\cal A}$.)
Then
$$H^0(\w{\cal A},\w{\cal L})=H^0({\cal A},{\cal L}^{\otimes k})
\otimes\left(\otimes_i \bar V_i\right).  \eqn\polyp$$
The multiplicity ${\rm mult}(V)$ of \micnic\
is therefore the same as the dimension of the $\widehat G$-invariant
subspace of $H^0(\w{\cal A},\w{\cal L})$:
$${\rm mult}(V)={\rm dim}\left(H^0(\w{\cal A},\w{\cal L})^{\w G}\right).
\eqn\jurfo$$

\REF\seshadri{V. B. Mehta and C. S. Seshadri,
``Moduli Of Vector Bundles On Curves With Parabolic Structures,''
Math. Ann. {\bf 248} (1980) 205.}
On the other hand, let $\w{\cal R}$ be the quotient of $\w{\cal A}$
by $\w G_{\bf C}$
(the quotient being taken in the sense of geometric invariant theory,
using the ample line bundle $\w{\cal L}$).  $\w{\cal R}$ is called
the moduli space of holomorphic bundles over $\Sigma$ with parabolic
structure, the parabolic structure being a reduction of the structure
group to $T$ at the marked points $x_i$.  (By a theorem of Mehta and
Seshadri [\seshadri],
$\w{\cal R}$ coincides with the moduli space of flat connections on
$P\to \Sigma\,\,
- \{x_i\}$ with certain branching about  the $x_i$, up to gauge
transformation.)
The $\w G_{\bf C}$-invariant line bundle $\w{\cal L}\to\w{\cal A}$
descends to a line bundle over $\w{\cal R}$, which we will also call
$\w{\cal L}$, whose sections over an open set $U\subset {\cal R}$
are $\widehat G$-invariant sections of $\w{\cal L}$ over the
inverse image of $U$ in $\w A$.  So in particular
$$H^0(\w{\cal R},\w{\cal L}) =H^0(\w{\cal A},\w{\cal L})^{\w G}.
\eqn\hurry$$
Both $\w{\cal R}$ and $\w{\cal L}$
depend on the $V_i$, but I will not indicate this in the notation.

The left hand side of \hurry\ is the space of non-abelian theta
functions with parabolic structure.
If we combine \micnic, \oxxo, \jurfo, and \hurry, we find
that the dimension of this space is naturally written as a correlation
function in the gauged WZW model:
$$\dim H^0(\w{\cal R},\w{\cal L})=\left\langle \prod_{i=1}^s\Tr_{V_i}g(x_i)
\right\rangle.\eqn\finalgo$$

\subsection{The Verlinde Algebra}

As a special case of this, the Verlinde algebra is defined as follows.
For given ``level'' $k$, the loop group of the compact Lie
group $G$ has a finite number of isomorphism classes of unitary,
integrable representations; their highest weights are a distinguished
list of isomorphism classes $V_\alpha,\,\,\alpha\in W$ of representations
of $G$.  Let $X$ be the ${\bf Z}$ module freely generated by the $V_\alpha$.
$X$ has a natural
metric given by $g(V_\alpha,V_\beta)=1$ if $V_\alpha=\bar V_\beta$
and otherwise $g(V_\alpha,V_\beta)=0$.
It also has a natural multiplication structure
that we will describe presently.
$X$ endowed with this structure is called the Verlinde algebra.

Using the metric on $X$, a multiplication
law $V_\alpha\cdot V_\beta=\sum_\gamma N_{\alpha\beta}{}^\gamma V_\gamma$
can be defined by giving a cubic form $N_{\alpha\beta\gamma}$ which
is interpreted as $\sum_\delta g_{\gamma\delta}N_{\alpha\beta}{}^{\delta}$.
Such a cubic form is defined as follows.

Take $\Sigma$ to be a curve of genus zero with three marked points
$x_i$, $i=1\dots 3$, labeled by integrable representations $V_{\alpha_i}$,
$\alpha_i\in W$.  The choice of the $\alpha_i$ and of a level $k$
determines a moduli space $\w{\cal R}$ of parabolic bundles
with a line bundle $\w{\cal L}$.  The structure constants of the Verlinde
algebra are
$$N_{\alpha_1,\alpha_2,\alpha_3}=\dim H^0(\w{\cal R},\w{\cal L}).
\eqn\yuyu$$
So in other words, from \finalgo, the Verlinde structure functions
are the genus zero three point functions of the $G/G$ model:
$$N_{\alpha_1,\alpha_2,\alpha_3}=\left
\langle\prod_{i=1}^3\Tr_{V_i}g(x_i)\right\rangle.
\eqn\josos$$

The basic phenomenon under study in the present paper is a relation
between the quantum cohomology of the Grassmannian and the $G/G$ model;
the result can be applied to the Verlinde algebra because of \josos.
The special case of a genus zero surface with three marked
points is fundamental because the general case can be reduced to this
by standard sewing and gluing arguments.
In fact, such sewing and gluing arguments, applied to
a genus zero curve with four marked points, yield the associativity
of the Verlinde algebra.

\subsection{Higher Cohomology}

Obviously, the above discussion has only a physical level of rigor.
Among many points that should be clarified I will single out one.

If the $V_i$ are integrable representations at level $k$, then the higher
cohomology $H^i(\w{\cal R},\w{\cal L}),\,\,i>0$ vanishes,
and $\dim H^0(\w{\cal R},\w{\cal L})$ coincides with the Euler characteristic
$\chi(\w{\cal R},\w{\cal L})=\sum_i(-1)^i\dim H^i(\w{\cal R},\w{\cal L})$.
{}From comments made to me by R. Bott and G. Segal, it appears that
for \finalgo\ to hold for arbitrary representations $V_i$
(perhaps not integrable),  one must replace $\dim H^0(\w{\cal R},\w{\cal L})$
by $\chi(\w{\cal R},\w{\cal L})$.  A rigorous treatment of the $G/G$
model should show the restriction to integrable representations
in deriving \finalgo; there may also be a supersymmetric version
of the derivation that naturally gives the Euler characteristic
and holds for all representations.

\section{Some Additional Properties}

The reader may wish at this stage to turn to \S3.  However, I will pause
here and in \S2.6 below
to explain a few additional facts that have their own interest
and will be needed at a few points in \S4.

\subsection{Topological Field Theory}

First of all, the gauged WZW theory of $G/H$ is in general
conformally invariant but not topologically invariant.  A
conformal structure appears in the definition of the Lagrangian.
However, for $H=G$ we have evaluated the
partition function of the $G/H$ model, and found it to
be an integer, independent of the conformal structure of $\Sigma$,
and equal to the dimension of the space of non-abelian theta functions.
This strongly suggests that the $G/G$ model is actually a topological
field theory.  Let us try to demonstrate that directly.

A conformal structure on $\Sigma$ can be specified by giving a metric
$h$, uniquely determined up to Weyl scaling.
Under a change in $h$, the change in the $G/G$ Lagrangian is
$$\delta L={k\over 8\pi}\int_\Sigma \d^2z\sqrt h (h^{z\bar z})^2
\left(\delta h_{\bar z\bar z}\Tr (g^{-1}D_zg)^2+\delta h_{zz}
\Tr (D_{\bar z} g\cdot g^{-1})^2\right).  \eqn\polo$$
Though this expression does not vanish identically, it vanishes
when the classical equations of motion are obeyed.  In fact,
under a variation of the connection $A$, the Lagrangian changes
by
$$\delta' L= {k\over 2\pi}\int_\Sigma d^2z\sqrt h h^{z\bar z}
\Tr\left(\delta A_{\bar z} g^{-1}D_zg-\delta A_z D_{\bar z}g\cdot g^{-1}
\right).        \eqn\moomoo$$
So the classical Euler-Lagrange equations, asserting
the vanishing of $\delta' L$, are
$$0 = g^{-1}D_z g= D_{\bar z}g \cdot g^{-1}. \eqn\roomoo$$
Since \polo\ vanishes when \roomoo\ does,
the $G/G$ model is classically a topological field theory.
Quantum mechanically
the analog of using the equations of motion is to make a suitable
change of variables in the path integral.   In this case, we consider
the infinitesimal redefinition of $A$
$$\eqalign{\delta A_z & = {1\over 4}\delta h_{zz}h^{z\bar z}D_{\bar z}g\cdot
              g^{-1} \cr
           \delta A_{\bar z} & = -{1\over 4} \delta h_{\bar z\bar z}
         h^{z\bar z} g^{-1}D_zg. \cr}         \eqn\ujmoo$$
(This is a complex change of coordinates that entails an infinitesimal
displacement of the integration contour in the complex plane, or
more exactly a displacement of the cycle of integration in the complexification
of ${\cal A}$.)
Substituting in \roomoo, we see that the
Lagrangian $L(g,A)$ is invariant under a change of metric on $\Sigma$
compensated by the transformation \ujmoo\ of the field variables.
The path integral for the partition function
$$ \int \D g\,\,\,\D A \,\,\,\exp(-L(A,g)) \eqn\jmoo$$
is therefore invariant
under the combined change of metric and integration variable,
provided the measure
$\D A$ is invariant.   To this effect, we must compute a Jacobian or,
at the infinitesimal level, the divergence
of the vector field that generates the change of variables \ujmoo.
This is formally
$$\int_\Sigma\left({\delta\over\delta A_z(x)}\delta A_z(x)
+{\delta\over\delta A_{\bar z}(x)}\delta A_{\bar z}(x).\right) \eqn\gmoo$$
This vanishes, as $\delta A_z$ is indepependent of $A_z$ and $\delta
A_{\bar z}$ is independent of $A_{\bar z}$.
This completes the explanation of why the $G/G$ model
is a topological field theory.

Let us note now that the other Euler-Lagrange equation of motion,
obtained by varying with respect to $g$, is
$$D_{\bar z}(g^{-1}D_zg) +F_{\bar zz}= 0, \eqn\cxxon$$
with $F$ the curvature of the connection $A$.  So given
\roomoo, this implies that
$$F=0. \eqn\czonzo$$

\subsection{Comparison To The Obvious Topological Field Theory}

If the goal were to construct a topological field theory using
the fields $g,A$, the more obvious way to do it would be to take
the Lagrangian to be simply
$$L'(g,A)=-ik\Gamma(g,A), \eqn\hixon$$
which manifestly corresponds to a topological field theory, since
it is defined without use of any metric or conformal structure.
How does this theory compare to the $G/G$ WZW model?

More generally, let us consider the family of theories
$$L_{k'}(g,A)=-{k'\over 8\pi}\int_\Sigma\Tr g^{-1}\d_Ag\wedge
*g^{-1}\d_Ag  -ik\Gamma(g,A), \eqn\cargo$$
with positive $k'$.
This coincides with the $G/G$ model at $k'=k$, and with the manifestly
topologically invariant model at $k'=0$.
It is straightforward to work out that the classical equations
of motion are
$$ 0 =g^{-1}D_zg -\lambda D_zg\cdot g^{-1}=D_{\bar z}g\cdot g^{-1}-
\lambda g^{-1}D_{\bar z}g, \eqn\cimbo$$
with
$$\lambda = {k'-k\over k'+k}.\eqn\argo$$
For $0<k'<\infty$, one has
$$-1<\lambda<1. \eqn\dgon$$
\cimbo\ implies
$$\d_Ag = 0 .\eqn\noki$$
For instance, the first equation in \cimbo\ is equivalent to
$$\left(1-\lambda {\rm Ad}(g)\right)(D_zg)=0, \eqn\ncon$$
with ${\rm Ad (g)}(x)=gxg^{-1}$.  Since $|{\rm Ad}(g)|\leq 1$ and $|\lambda|
<1$, \ncon\ implies $D_zg=0$, and similarly \cimbo\ implies $D_{\bar z}g=0$.

Given that \noki\ follows from the classical equations of motion,
the same sort of reasoning as above shows that the Lagrangians
$L_{k'}$ describe a family of topological field theories: a change
of metric can be compensated by a change of integration variable
with trivial Jacobian.

Now, to study the $k'$ dependence, look at
$${\partial L_{k'}\over \partial k'}=-{1\over 8\pi}\int_\Sigma
\Tr (g^{-1}\d_Ag\wedge *g^{-1}\d_Ag).           \eqn\xoxo$$
By virtue of
\noki, this expression vanishes by the classical equations of motion,
so classically the family of theories governed by $L_{k'}$ is
constant.

{}From the above discussion, we know how we should proceed quantum
mechanically:
we should find a change of integration variable that compensates
for the $k'$ dependence of the Lagrangian.  Such a change
of variable exists because of \noki; one can take explicitly
$$\delta A_z=-{\delta k'\over k+k'}\left(1-\lambda{\rm Ad}(g^{-1})\right)^{-1}
 (D_zg\cdot g^{-1}).\eqn\coconon$$
Now, however, a difference arises from our earlier discussion. Because
$\delta A_z$ is a function of $A_z$, the Jacobian of the transformation
in \coconon\ is not necessarily 1; the integration measure in the path
integral may not be invariant.
The change of the integration measure is formally
$$\int_\Sigma \Tr {\delta\over\delta A_z(x)}\delta A_z(x).
\eqn\cucuin$$
Since
$${\delta \over\delta A_z(x)}\delta A_z(y)\sim \delta^2(x,y),
      \eqn\ucuin$$
this is ill-defined, proportional to $\delta^2(0)$.  In any
event, since \cucuin\ is the integral over $\Sigma$ of a local
quantity, any regularization should be of that form.  Quantities
analogous to \cucuin\ are regularized (albeit in a slightly
{\it ad hoc} fashion) in [\blau], in deriving eqn. (6.22).
I will not repeat such a calculation here, but I will just explain
what general form the answer must have,  by asking what is
the most general possible perturbation of the $G/G$ model.

\subsection{The Complete Family Of Theories}

\REF\maybe{S. Elitzur, A. Forge, and E. Rabinovici,
``On Effective Theories Of Topological Strings,''
Nucl. Phys. {\bf B388} (1992) 131.}
Let us simply go back to the gauged WZW model of $G/G$, and
ask what kind of perturbations it has (see also [\maybe]).
We permit the perturbation
of the Lagrangian to be the integral of an arbitrary local functional
of $g,A$, and a metric $h$ on $\Sigma$.
In this way we will obtain continuous perturbations of the
$G/G$ model, but forbid discrete perturbations (notably changes in $k$)
that cannot be described via the addition of a local functional
to the Lagrangian.

Perturbations that vanish
by the classical equations of motion are irrelevant,  since they
can be eliminated by a change of integration variables as described
above.  (Even if the integration measure is not invariant under
the change of variables, changes of variables can be used to eliminate
the perturbations that vanish by the equations of motion in favor
of other perturbations that do not so vanish.)
In classifying perturbations, we therefore can work modulo operators
that vanish by the classical equations of motion.  Given \roomoo\
and \czonzo, this means that we can discard anything proportional to
$\d_Ag$ or $F$.

The gauge invariant local operators, modulo operators that vanish
by the equations of motion, are generated by operators of the form
$U(g)$, with $U$ some function on $G$ that is invariant under conjugation.
Since $U(g)$ is a zero-form, to construct from it a perturbation
of the Lagrangian, we need also a metric $h$ on $\Sigma$, or at least
a measure $\mu$, such as the Riemannian measure.  The curvature scalar
of $h$ will be called $R$.  The most interesting
perturbations are
$$Q_U= \int_\Sigma \d \mu \,\,U(g) \eqn\nxon$$
and
$$S_U= \int_\Sigma d^2z \sqrt h  R \,  \, U(g). \eqn\xon$$

\nxon\ breaks the diffeomorphism invariance of the $G/G$ model
down to invariance under the group of diffeomorphisms that preserve
the measure $\mu$.  The $G/G$ model perturbed as in
\nxon\ is an interesting family of theories invariant under
area-preserving diffeomorphisms (and reducing for $k\to\infty$ to
two dimensional Yang-Mills theory, which has the same invariance).

Slightly less obviously, the $G/G$ model perturbed by \xon\ is still
a topological field theory.  In fact, under an infinitesimal change in $h$,
$\sqrt h R$ changes by a total derivative (so that
$\int_\Sigma \d^2z\sqrt h R $ is a topological invariant, a multiple
of the Euler characteristic).   After integrating by parts,
the change in \xon\ under a change in $h$ is
$$\delta S_U\sim \int_\Sigma\d^2z \sqrt h \left(
\delta h_{i'j'}-h_{i'j'}h^{kl}\delta h_{kl}\right) h^{i'i}h^{j'j}
D_iD_j U(g).  \eqn\uponon$$
This vanishes by the equations of motion, since
$\d_A(g)=0$ implies $\d U=0$.
Hence one can compensate for $\delta S_U$ with a redefinition of
$A$ (and the Jacobian for the transformation is trivial, since
the requisite $\delta A$ is independent of $A$).

Other perturbations, such as $\int_\Sigma \d^2z \sqrt h R^2 U(g)$,
are less interesting, since (i) they do not possess the large
invariances of the theories perturbed by $Q_U$ or $S_U$; (ii)
they vanish as a negative power of $t$ if the metric of $\Sigma$
is scaled up by $h\to t h$, $t>>1$.  The latter property means
that in most applications of these systems, such perturbations
(if not prevented by (i)) can be conveniently eliminated.

Since the most general perturbation of the $G/G$ model that
preserves the diffeomorphism invariance is of the form of $S_U$,
the regularized version of \cucuin\ must be
equivalent to $S_U$ for some $U$.  By the same token,
for any $k'$, the $G/G$ model
must be equivalent to the $L_{k'}$ model perturbed by some $S_U $
(with a $k'$-dependent $U$), and vice-versa.  In particular,
setting $k'=0$, the $G/G$ model is equivalent to the
manifestly topologically invariant model with Lagrangian
$-ik\Gamma(g,A)$, perturbed by some $S_U$. The requisite $ U$'s in these
statements can in fact be computed at least heuristically
along the lines of the derivation of eqn. (6.22) of [\blau],
but I will not do so here.

\subsection{Interpretation}

Note that the conjugation-invariant function $U(g)$ that entered
above can be expressed as a linear combination of the characters
$\Tr_Vg$, as $V$ runs over irreducible representations of $G$.
These are precisely the operators whose correlation functions
were interpreted algebro-geometrically in \finalgo, so the
theories obtained by perturbing the $G/G$ model are all computable
in terms of the Verlinde algebra.

\section{Abelianization}

I will now briefly describe another interesting facet of the $G/G$
model, introduced in [\blau], which apart from its beauty will enter
at a judicious moment in \S4.
\foot{A computation reaching a rather similar conclusion is sketched
in [\gerasimov], but unfortunately the fermionic symmetry
$\delta $ introduced in equations (71)-(74) of that paper does not obey
$\delta^2=0$, which would be needed to justify the computation.
I will therefore concentrate on sketching the argument of [\blau].}

A recurring and significant theme in the theory of compact
Lie groups is the reduction to the maximal torus $T$, extended
by the Weyl group $W$.  As explained
in [\blau], the $G/G$ model admits such an reduction to the
maximal torus.  It is equivalent to the $T/T$ model
(that is, the $G/H$ model with both $G$ and $H$ set equal to $T$)
perturbed by $S_U$, where $U$ is a certain Weyl-invariant function
on $T$ and $S_U$ is defined in \xon.

At the level of precision explained in [\blau], the abelianization
of the model proceeds as follows.  Pick a maximal torus $T\subset G$,
with Lie algebra ${\bf t}$.
Impose the ``gauge condition'' $g\in T$.
\foot{This is not really valid globally as a gauge condition.
One must think in terms of integrating over the fibers of the map
$G\to T/W$ that maps a group element to its conjugacy class.}
Decompose the connection
as $A=A_0+A_\perp$, where $A_0$ is the part of the connection valued
in ${\bf t}$, and $A_\perp$ is valued in the orthocomplement $\bf t_\perp$
of ${\bf t}$.
In this gauge the $G/G$ Lagrangian takes the form
$$L_{G/G}(g,A)=L_{T/T}(g,A_0) -{k\over 2\pi}\int_\Sigma \d^2z\Tr \left(
A_{\perp,z}A_{\perp,\bar z}-A_{\perp,z}g A_{\perp,\bar z}g^{-1}\right).
            \eqn\ombo$$
Here
$$L_{T/T}(g,A_0)=kI_{T/T}(g,A_0) \eqn\rombo$$
is the Lagrangian of the $T/T$ model, at level $k$.
The $G/G$ model, in this gauge, differs from the $T/T$ model
by the last term in \ombo, which involves $A_\perp$.
To reduce the $G/G$ model to something like the $T/T$ model,
one must ``integrate out'' $A_\perp$ to reduce to a description
involving $g$ and $A_0$ only.  Happily, the $A_\perp$ integral is
Gaussian:
$$\int \D A_\perp\exp\left({k\over 2\pi}\int_\Sigma \d^2z\Tr\left(
A_{\perp,z} A_{\perp,\bar z}-A_{\perp,z}gA_{\perp,\bar z}g^{-1}\right)\right).
\eqn\umco$$
Such a Gaussian integral formally gives rise to a determinant
(as we briefly explain in \S3.5 below).  In comparing the
$G/G$ model to the $T/T$ model, another determinant arises:
the Fadde'ev-Popov determinant comparing the volume of $\w G$ to
the volume of $\w T$.  These two determinants are rather singular
but at the same time extremely simple, because the exponent in
\umco\ (like the corresponding expression in the Fadde'ev-Popov determinant)
is a local functional without derivatives.
In [\blau], Blau and Thompson calculate these determinants, with
a plausible regularization, and argue that the $G/G$ model is
equivalent to a $T/T$ model with Lagrangian
$$\w L_{T/T}(g,A_0)=(k+\rho)I(g, A_0)-{1\over 4
\pi}\int_\Sigma
\d^2x \sqrt h R \log\det{}_{{\bf t}_\perp}(1-{\rm Ad}(g)). \eqn\polyp$$
Here $\rho$ is the dual Coxeter number of $G$, and $\det_{{\bf t}_\perp}
(1-{\rm Ad}(g))$ is the determinant of $1-{\rm Ad}(g)$, regarded
as an operator on ${\bf t}_\perp$.  (This well-known Weyl-invariant function
enters in the Weyl character formula, where it has a somewhat
similar origin,
involving a comparison of the volumes of $G$ and $T$.)
In \S4.6, we will have occasion to use \polyp\ for the case that
$G=U(k)$.  For that case, if $g={\rm diag}(\sigma_1,\dots,\sigma_k)$,
the eigenvalues of $1-{\rm Ad}(g)$ acting on ${\bf t}_\perp$
are the numbers $1-\sigma_i\sigma_j{}^{-1}$,
for $1\leq i,j\leq k$, $i\not= j$.  Hence the correction term in \polyp\
becomes in this case
$$\Delta L =-{1\over 4\pi}\int_\Sigma d^2x \sqrt h R \left(\sum_{i\not= j}
\ln(\sigma_i-\sigma_j)-(k-1)\sum_i\ln\sigma_i\right). \eqn\lateruse$$

Given the role of the $G/G$ model in counting non-abelian theta
functions, its reduction to a $T/T$ model is a
kind of abelianization of the problem of counting such functions.
In \S7.3 of [\blau], this is pursued further to obtain a completely
explicit count of non-abelian theta functions for $G=SU(2)$.
The role of the endpoint contributions in equation (7.14) of that
paper still deserves closer study.

\chapter{The Quantum Cohomology Of The Grassmannian}

\REF\frenkel{I. B. Frenkel, ``Representations of affine Lie algebras,
Hecke modular forms and Korteweg-de Vries equations,'' in
{\it Lie Algebras And Related Topics},  Lecture Notes In Mathematics
Vol. 933 (Springer-Verlag, 1982), p. 71; ``Representations Of Kac-Moody
Algebras And Dual Resonance Algebras,'' in {\it Lectures In
Applied Mathematics}, vol. 21 (American Mathematical Society, 1985), p.  325.}
\REF\segal{A. Pressley and G. Segal, {\it Loop Groups} (Oxford University
Press, 1986).}
\REF\schnitzer{E. J. Mlawer, S. G. Naculich, H. A. Riggs, and H. J.
Schnitzer, ``Group Level Duality of WZW Fusion Coefficients And
Chern-Simons Link Observables,'' Nucl. Phys. {\bf B352} (1991) 863;
S. G. Naculich, H. A. Riggs, and H. J. Schnitzer,
``Group Level Duality In WZW Models And Chern-Simons Theory,''
Phys. Lett. {\bf B246} (1990) 417.}
\REF\naka{T. Nakanishi and A. Tsuchiya, ``Level Rank Duality Of WZW Models
In Conformal Field Theory,'' Commun. Math. Phys. {\bf 144} (1992) 351.}
The Grassmannian $G(k,N)$ is the space of all $k$ dimensional
subspaces of a fixed $N$ dimensional complex vector space $V\cong
\C^N$.  If we want to make the dependence on $V$ explicit, we write
$G_V(k,N)$.

By associating with a $k$ dimensional subspace of $ V$
the $N-k$ dimensional orthogonal subspace of the dual space $V^*$, we see that
$G_V(k,N)\cong G_{V^*}(N-k,N)$.  The relation that we will explain
here and in \S4 between the
Verlinde algebra of $U(k)$ at level $(N-k,N)$\foot{That is,
at levels $N-k$ and $N$ for the $su(k)$ and $u(1)$ factors
in the Lie algebra of $U(k)$.} and the quantum cohomology of
$G(k,N)$ therefore implies that the Verlinde algebra
of $U(k)$ at level $(N-k,N)$ coincides with that of $U(N-k)$
at level $(k,N)$.  This is a surprising fact that had been
noted earlier. (For instance, see [\frenkel] and [\segal, p. 212, Proposition
(10.6.4)] for $k\leftrightarrow N-k$ symmetry of loop group representations
and [\schnitzer,\naka] for such symmetry of the Verlinde algebra.)

One way to describe $G(k,N)$ is as follows.  Let $B$ be the space of all
linearly independent $k$-plets $e_1,\dots,e_k\subset V$.  A point in $B$
labels a $k$-plane $V$ with a basis.  The group $GL(k,\C)$ acts on $B$ by
change of basis, $e_i\to \sum_j W_i{}^je_j,\,\,\,W\in GL(k,\C)$.  Since
$GL(k,\C)$ acts simply transitively on the space of bases of $V$, upon
dividing by $GL(k,\C)$ we precisely forget the basis and therefore
$$G(k,N)=B/GL(k,\C). \eqn\furtful$$

$B$ is dense and open in the $k$-fold product
$\C^{kN}=V\times V\times \dots \times V$
(since the generic $k$-plet $e_1,\dots , e_k\subset V$ is a basis
of $V$), so $G(k,N)$ is a quotient of a dense open subset of $\C^{kN}$
by $GL(k,\C)$.  In fact, $G(k,N)$ is the good quotient of $\C^{kN}$
by $GL(k,\C)$ that would be constructed in geometric invariant theory.

There is also a symplectic version of this, which will be more relevant
in what follows.  Pick a Hermitian metric on $V$ so that $V^k=\C^{kN}$
gets a metric and a symplectic structure.  In linear coordinates
$\phi^{is}$, $i=1\dots k$, $s=1\dots N$ on $\C^{kN}$, the symplectic
form is
$$\omega=i\sum_{i,s}\d\phi^{is}\wedge \d \bar\phi_{is}.\eqn\sympst$$
$\omega$ is not invariant under $GL(k,\C)$, but it is invariant under
a maximal compact subgroup $U(k)\subset GL(k,\C)$.

To this symplectic action is associated a ``moment map'' $\mu$
from $\C^{kN}$ to the dual of the Lie algebra of $U(k)$, given by the
angular momentum functions that generate $U(k)$ via Poisson brackets.
In this case we can take the moment map to be
$$\mu:(e_1,\dots,e_k)\rightarrow \{(e_i,e_j)-\delta_{ij}\}.\eqn\morfo$$
In other words, $\mu=0$ precisely if the vectors $e_1,\dots, e_k$
are orthonormal.

Every $k$-plane has an orthonormal basis, unique up to the action of
$U(k)$, so
$$G(k,N)=\mu^{-1}(0)/U(k). \eqn\tuggo$$
This is the description of $G(k,N)$ that we will actually use.
We will also want to remember one fact: $\mu$ is a quadratic function
on the real vector space underlying $\C^{kN}$.  In components,
$$\mu^i{}_j=\sum_s\phi^{is}\bar\phi_{js}-\delta^i{}_j.\eqn\tuffo$$

\section{Cohomology}

Now we need to discuss the cohomology of $G(k,N)$.  We begin with the
classical cohomology.  Over $G(k,N)$ there is a ``tautological''
$k$-plane bundle $E$ (whose fiber over $x\in G(k,N)$ is the $k$
plane in $V$ labeled by $x$) and a complementary bundle $F$
(of rank $N-k$):
$$0\to E\to V\cong \C^N\to F \to 0. \eqn\compbun$$
Obvious cohomology classes of $G(k,N)$ come from Chern classes.
We set
$$ x_i=c_i(E^*), \eqn\mipp$$
where $*$ denotes the dual.  (It is conventional to use $E^*$ rather
than $E$, because $\det E^*$ is ample.)
This is practically where Chern classes come from, as $G(k,N)$ for
$N\to \infty$ is the classifying space of the group $U(k)$.
It is known that the $x_i$ generate $H^*(G(k,N))$ with certain relations.
The relations come naturally from the existence of the complementary
bundle $F$ in \compbun.  Let $y_j=c_j(F^*)$, and let
$c_t(\cdot)=1+tc_1(\cdot)+t^2c_2(\cdot)+\dots$.  Then
as a consequence of \compbun,
$$ c_t(E^*)c_t(F^*)=1, \eqn\ombun$$
and $H^*(G(k,N))$ is generated by the $x_i,y_j$ with relations
\ombun.
If one wishes,
these relations can be partially solved to express the $y_j$ in terms
of the $x_i$ (or vice-versa).

\REF\instantons{M. Dine, N. Seiberg, X.-G. Wen, and E. Witten,
``Non-Perturbagtive Effects On The String World Sheet, I, II''
Nucl. Phys. {\bf B278} (1986) 769, {\bf B289} (1987) 319.}
\REF\gromov{M. Gromov, ``Pseudo-Holomorphic Curves In Symplectic Manifolds,''
Invent. Math. {\bf 82} (1985) 307.}
Now we come to the quantum cohomology, which originally entered in
string theory, where [\instantons]
it enters the theory of the Yukawa couplings
(which are related to quark and lepton masses), and in Floer/Gromov
theory of symplectic manifolds [\gromov].  Additively, the quantum cohomology
is the same as the classical one, but the ring structure is different.

Giving a ring structure on $W=H^*(G(k,N))$ is the same as giving the
identity $1\in W$ and a cubic form
$$(\alpha,\beta,\gamma)=\int_{G(k,N)}\alpha\cup\beta\cup\gamma.\eqn\wurko$$
The cubic form determines a metric
$$g(\alpha,\beta)=(\alpha,\beta,1), \eqn\bomxo$$
and given a metric the cubic form $W\times W\times W\to \C$
determines a ring structure $W\times W\to W$.

So I will explain the quantum cohomology ring by describing the quantum
cubic form.  To this aim, let $\Sigma$ be a closed oriented two-manifold
(which in string theory would be the ``world-sheet,'' analogous to the
world-line of a particle).  Let $P\in \Sigma$.  Let
${\cal W}={\rm Maps}(\Sigma,G(k,N))$.  Evaluation at $P$ gives
a map
$${\cal W}\underarrow{{\rm ev}(P)} G(k,N), \eqn\evmap$$
by which $\alpha\in H^*(G(k,N))$ pulls back to $\widehat\alpha(P)
={\rm ev}(P)^*(\alpha)\in H^*({\cal W})$.

Now pick a complex structure on $\Sigma$, and let ${\cal M}\subset
{\cal W}$ be the space of holomorphic maps of $\Sigma$ to $G(k,N)$.
We have ${\cal M}=\cup_\lambda {\cal M}_\lambda$, with ${\cal M}_\lambda$
being the connected components of ${\cal M}$.
In the case
of the Grassmannian, the components ${\cal M}_
\lambda$ are determined by the degree, defined as follows.
If $\eta=c_1(E^*)$, which generates $H^2(G(k,N),{\bf Z})$,
and $\Phi:\Sigma\to G(k,N)$ is such that
$\int_\Sigma\Phi^*(\eta)=d$, then $\Phi$ is said to be of degree $d$.
Since $\det E^*$ is ample, holomorphic curves only exist for $d\geq 0$.

The quantum cubic  form
is defined as follows (ignoring analytical details and tacitly assuming that
the ${\cal M}_\lambda$ are smooth and compact).
Let $\Sigma$ be of genus zero.  Let $P,Q,R$
be three points in $\Sigma$.  Then for $\alpha,\beta,\gamma\in H^*(G(k,N))$,
we set
$$\langle \alpha,\beta,\gamma\rangle=\sum_d  e^{-dr}\cdot \int_{{\cal M}_d}
\widehat\alpha(P)\cup \widehat\beta(Q)\cup \widehat\gamma(R),
\eqn\tangle$$
with $r$ a real parameter.

In what sense does $\langle\alpha,\beta,\gamma\rangle$ generalize the
classical cubic form?  One component of ${\cal M}$, namely ${\cal M}_0$,
consists of constant maps $\Sigma\to G(k,N)$.  This component is a copy
of $G(k,N)$ itself.  Under that identification the evaluation maps
at $P,Q$, and $R$ all coincide with the identity, so the contribution
of ${\cal M}_0$ to $\langle\alpha,\beta,\gamma\rangle$ coincides with
the classical cubic form defined as in \wurko.  The quantum cubic
form differs from the classical one by contributions of the rational
curves of higher degree.  These contributions are small for $r>>0$.
In practice, for dimensional reasons, for every given $\alpha,\beta,\gamma$
of definite dimension, the sum in \tangle\ receives a contribution from
at most one value of $d$.  (This is in marked contrast to the much-studied
case of a Kahler manifold of $c_1=0$, where every positive $d$ can contribute
to the same correlation function.)  Therefore, no information is lost
if we set $r=0$, and that is what we will do in the rest of this section.

It follows from the definition (for any Kahler manifold, not just
the Grassmannian) that
$$\langle \alpha,\beta,1\rangle =(\alpha,\beta,1) \eqn\nurgo$$
and thus that the classical and quantum metrics coincide.
This is equivalent to the statement that rational maps of positive
degree do not contribute to $\langle\alpha,\beta,1\rangle$.
In fact (as $\widehat 1(R)=1$),
the contribution of a component ${\cal M}_\lambda$
of positive degree is
$$\int_{\cal M_\lambda}\widehat\alpha(P)\cup\widehat\beta(Q).
\eqn\ongo$$
A group $F\cong\bf C^*$
acts on ${\bf CP}^1$ leaving fixed the points $P$ and $Q$.
$F$ acts freely on ${\cal M}_\lambda$, if ${\cal M}_\lambda$ is
a component of rational maps of positive degree.  The classes
$\widehat\alpha(P)$ and $\widehat\beta(Q)$ in the cohomology
of ${\cal M}_\lambda$ are pullbacks from ${\cal M}_\lambda/F$.
Therefore, on dimensional grounds \ongo\ vanishes.

\section{The Grassmannian}

\def\Z{{\bf Z}}
Let us now  work out the quantum cohomology ring of the Grassmannian.
As a preliminary, we note that the contribution of a moduli
space ${\cal M}_d$
to the quantum cubic form obeys an obvious dimensional condition:
it vanishes unless the sum of the dimensions of $\alpha,\beta,\gamma$
equals the (real) dimension of ${\cal M}_d$.
The component ${\cal M}_d$ of genus zero holomorphic curves of
degree $d$ in $G(k,N)$ has (according to the Riemann-Roch theorem)
complex dimension $\dim_{{\bf C}} G(k,N)+dN$. The fact that this
depends on $d$ means that the dimensional condition depends on $d$
and therefore that the quantum cohomology ring is not $\Z$-graded.
However, the fact that the real dimensions are all equal modulo
$2N$ means that the cohomology is $\Z/2N\Z$-graded.

\REF\yaulect{E. Witten, ``Two Dimensional Gravity And Intersection
Theory On Moduli Space,'' Surveys in Differential Geometry
{\bf 1} (1991) 243.}
Returning to the relations $c_t(E^*)c_t(F^*)=1$ that define the cohomology
of the Grassmannian, we see that (as the left hand side is  {\it a priori}
a polynomial in $t$ of degree $N$) the classical relations are of dimension
$0,2,4,\dots, 2N$.  To a classical relation of degree $2k$, the rational
curves of degree $d>0$ will  add a correction of degree $2k-2dN$; this
therefore must vanish unless $k=N$ and $d=1$.  Therefore, of the defining
relations of the cohomology, the only one subject to a quantum correction
is the ``top'' relation $c_k(E^*)c_{N-k}(F^*)=0$, and the correction is
an element of $H^*(G(k,N))$
of degree 0 and hence simply an integer.  So the non-trivial effect of
the quantum corrections will be simply to generate a relation of the form
$$ c_k(E^*)c_{N-k}(F^*)= a, \eqn\rellform$$
for some $a\in \Z$.
Moreover, $a$ is to be computed by examining rational curves in the
Grassmannian of degree 1.  We will find that $a=(-1)^{N-k}$, so
the quantum cohomology ring can be described by the relations
$$c_t(E^*)c_t(F^*)=1+ (-1)^{N-k}t^N. \eqn\ellform$$

This correction has been described previously [\yaulect] in the special case
of $k=1$ (complex projective space).
Despite its simple form, the correction has a dramatic effect:
while the classical cohomology ring is nilpotent (in the sense that every
element of positive degree is nilpotent), the quantum cohomology ring
is semi-simple.  This is evident in its Landau-Ginzburg description
[\lerche,\intrilligator,\gepner] which we consider presently.

\subsection{Computation Of $a$}

For $X$ a submanifold of $G(k,N)$, let $[X]$ be its Poincar\'e dual
cohomology class.  For instance, for $p$ a point in the Grassmannian,
$[p]$ is a top dimensional class, obeying $g(1,[p])=1$.  (It does
not matter here if the metric $g(~,~)$ is defined using the classical
or quantum cubic form, since we have seen that
these determine the same metric.)
The definition of the quantum ring structure from the quantum cubic
form is such that $a=c_k(E^*)c_{N-k}(F^*)$ can be computed
as
$$ a =\langle c_k(E^*),c_{N-k}(F^*),[p]\rangle. \eqn\ormo$$

$c_k(E^*)$ equals the Poincar\'e dual of the zero locus
of a generic section of $E^*$.  The dual of the exact sequence
\compbun\ reads
$$ 0 \to F^*\to V^* \to E^* \to 0, \eqn\yiro$$
with $V^*$ a fixed $N$ dimensional complex vector space.  The image
in $E^*$ of any fixed vector $w\in V^*$ gives a holomorphic section $\bar w$
of $E^*$.  If as before $e_1,\dots ,e_N$ is a basis of $V$, and
$w$ is the linear form that maps $\sum_{i=1}^N r^ie_i$ to $r^1$,
then the restriction of $w$ to $E\subset V$ vanishes precisely if
$E$ consists only of vectors with $r^1=0$.  This is a copy of $G(k,N-1)$
which we will call $X_w$.  Since $\bar w$ has only a simple zero along
$X_w$ (any $E$ can be perturbed in first order to get one for which
$\bar w\not= 0$), we have
$$c_k(E^*)=[X_w].\eqn\juniper$$

For future use, let us note that
$$\int_{G(k,N)}c_k(E^*)^{N-k} = 1 .   \eqn\conc$$
Indeed, we can pick $N-k$ holomorphic sections of $E^*$ whose
zero sets intersect transversely at a single point.  To do so,
let $w_i$ for $i=1,\dots, N-k$ be the linear form on $V$
that maps $\sum_{i=1}^Nr^ie_i$ to $r^i$.  Then the $\bar w_i$
have the required properties, vanishing precisely for $E$ the
$k$-plane spanned by $e_{N-k+1},\dots, e_N$.

Now let us compute $c_{N-k}(F^*)=(-1)^{N-k}c_{N-k}(F)$.  Under the
holomorphic surjection $V\to F$, any vector $v\in V$ projects
to a holomorphic section $\bar v$ of $F$.  $\bar v$ vanishes precisely
if $v\in E$; let $Y_v=\{E\in G(k,N)|v\in E\}$.  Then $\bar v$ has a simple
zero along $Y_v$, so $c_{N-k}(F)=[Y_v]$ and therefore
$$c_{N-k}(F^*) = (-1)^{N-k} [Y_v]. \eqn\berry$$

Rational curves of degree one in $G(k,N)$
can all be described as follows.  Let $(s,t)$ be
homogeneous coordinates for ${\bf CP}^1$.  For $r_1,\dots,
r_k$ a set of $k$ linearly independent vectors in the $N$ dimensional
vector space $V$, let $\{r_1,\dots,r_k\}$ be the $k$-plane
that they span.  Then a rational curve of degree one in $G(k,N)$ is of the
form
$$(s,t)\to \{sr_0+tr_1,r_2,r_3,\dots, r_{k}\}, \eqn\witho$$
with $r_0,\dots, r_{k}$ being linearly independent vectors in $V$.

We have to calculate
$$a=(-1)^{N-k}\int_{{\cal M}_1}\widehat{[X_w]}(P)\cup \widehat{[Y_v]}(Q)\cup
\widehat{[p]}(R). \eqn\hdndn$$
Here ${\cal M}_1$ is the space of degree 1 rational curves,
$w\in V^*$, $v\in V$, and $P,Q,R$ are points in ${\bf CP}^1$.
If everything is sufficiently generic, $a$ is simply the number of degree
one curves that pass through $X_e$ at $P$, through $Y_f$ at $Q$, and
through $p$ at $R$.

We choose $p$ to be an arbitrary point in $G(k,N)$ corresponding
to a $k$-plane spanned by vectors $v_1,\dots ,v_k$.
We take $v=v_0$ to be linearly independent of these, and
we pick $w$ to be any linear form that maps $v_0$ to 1, $v_1$ to $-1$,
and the $v_j$ of $j>1$ to 0.

{}From the explicit description of degree one curves in \witho, we see that
the $k$-planes represented by points in the image of such a curve
are subspaces of a common $k+1$-plane.  For a curve that passes
through $Y_v$ at $Q$ and through $p$ at $R$, this is clearly
the $k+1$-plane $W$ spanned by $v_0,v_1,\dots ,v_k$.
Requiring that the curve pass also through $X_w$ at $P$
determines the curve uniquely.  For instance, if $Q=(1,0)$,
$R=(0,1)$, and $P=(1,1)$, then the degree 1 curve must be
$$(s,t)\to \{v_0s+v_1t,v_2,\dots,v_k\}. \eqn\lovon$$
The subvarieties $\widehat {[X_w]}(P)$, $\widehat {[Y_f]}(Q)$,
and $\widehat {[p]}(R)$ of $\cal M_1$ meet transversely at that point,
so we get finally
$$a=(-1)^{N-k} \eqn\humpback$$
as claimed above.

\subsection{Landau-Ginzburg Formulation}

Write
$$c_t(E^*)=\sum_{i=0}^kx_it^i, \eqn\jurry$$
with $x_i=c_i(E^*)$.  Define functions $y_j(x_i), \,\,j\geq 0$ by
$$ {1\over c_t(E^*)}=\sum_{j\geq 0}y_j t^j. \eqn\purry$$
Classically, the cohomology ring of $G(k,N)$ is described by the
relations
$$ y_j=0, \,\,{\rm for}\,\,N-k+1\leq j\leq N. \eqn\riflo$$

Let
$$-{\rm log} c_t(E^*)=\sum_{r\geq 0}U_r(x_1,\dots,x_k)t^r. \eqn\ulff$$
So
$$-t^jc_t(E^*)^{-1}=-{\partial\over\partial x_j}{\rm log}c_t(E^*)
=\sum_{r\geq 0}{\partial U_r\over\partial x_j}t^r. \eqn\hulff$$
Hence if
$$ W_0=(-1)^{N+1}U_{N+1} \eqn\pullf$$
then
$${\partial W_0\over\partial x_i} = (-1)^Ny_{N+1-i},\,\,{\rm for}\,\,
1\leq i\leq k. \eqn\bulff$$
So the defining relations of the classical cohomology take the form
$$ d W_0 = 0. \eqn\iflo$$

To obtain in a similar way the quantum cohomology ring, set
$$ W = W_0 +(-1)^kx_1. \eqn\xiflo$$
The relations $dW=0$ now give
$$y_{N+1-i}+(-1)^{N-k}\delta_{i,1}=0. \eqn\hugiflo$$
Therefore  the relation $c_t(E^*)\cdot(\sum_jy_jt^j)=1$ becomes
$$ \left(\sum_{i=0}^kx_it^i\right)\cdot\left(\sum_{j=0}^{N-k}y_jt^j
-(-1)^{N-k}t^N+O(t^{N+1})\right)=1. \eqn\cugiflo$$
Keeping only the terms of order at most $t^N$, this becomes
$$\left(\sum_{i=0}^kx_it^i\right)\cdot\left(\sum_{j=0}^{N-k}y_jt^j\right)=
 1 +(-1)^{N-k}t^N.  \eqn\pubiflo$$
This coincides with the quantum cohomology ring as described in \ellform.
The function $W$ is called the Landau-Ginzburg potential.

If we introduce the roots of the Chern polynomial
$$c_t(E^*)=\prod_{i=1}^k(1+\lambda_it), \eqn\oddball$$
then  $W$ can be written
$$ W(\lambda_1,\dots,\lambda_k)={1\over k+1}\sum_{j=1}^k\left(
\lambda_j{}^{N+1} +(-1)^k\lambda_j\right).  \eqn\graddy$$

Now let us discuss integration.
Integration defines a linear functional on the top dimensional cohomology
of $G(k,N)$, which is the cohomology in real dimension $2k(N-k)$:
$$f\to I( f)= \int_{G(k,N)}f.         \eqn\icco$$
Since $H^{2k(N-k)}(G(k,N))$ is one dimensional, any two linear
functionals on that space are proportional.  Such a linear functional
can be obtained as follows in the Landau-Ginzburg description.
We examine the classical case first.
If we consider $x_i=c_i(E^*)$ to be of degree $i$, then the top
dimensional cohomology consists of polynomials $f$ of degree $k(N-k)$
modulo the ideal generated by $\partial W_0/\partial x_i,\,\,\,i=1\dots k$.
Consider the linear form on homogeneous polynomials of degree $k(N-k)$
defined by
$$ J(f)= (-1)^{k(k-1)/2}\left({1\over 2\pi i}\right)^k\oint dx_1
\dots dx_k { f\over \prod_{i=1}^k \partial W_0/\partial x_i}. \eqn\hudd$$
The integration contour is a product of circles enclosing the poles
in the denominator.  $J(f)$ annihilates the ideal generated by $dW_0$,
since if $f$ is divisible by, say, $\partial W/\partial x_i$, then
one of the denominators in \hudd\ is canceled and one of the contour
integrals vanishes.

For $f$ of degree $k(N-k)$, the integral in \hudd\ is unaffected
if $W_0$ is replaced by $W$; this follows from taking the contour
integral on \hudd\ to be a large contour.  The integral can then
be evaluated as a simple sum of residues:
$$J(f)=(-1)^{k(k-1)/2}
\sum_{dW=0}{f\over \det\left({\partial^2
W\over\partial x_i\partial x_j}\right)}.
\eqn\werewolf$$
It is convenient to change variables from the $x_i$ to the $\lambda_a$.
One has
$$\left.
\det\left({\partial^2W\over\partial x_i\partial x_j}\right)\right|_{dW=0}
=\det\left({\partial^2W\over\partial \lambda_a\partial \lambda_b}
\right)\cdot\det\left({\partial\lambda_a\over\partial x_i}\right)^2.
\eqn\erewor$$
The Jacobian in the change of variables from $x_i$ to $\lambda_a$ is
the Vandermonde determinant:
$$\det\left({\partial x_i\over\partial\lambda_a}\right)^2
=\prod_{a<b}(\lambda_a-\lambda_b)^2.  \eqn\jerwor$$
So
$$J(f)={(-1)^{k(k-1)/2}\over k!}\sum_{dW(\lambda_a)=0}
{f\cdot\prod_{a<b}(\lambda_a-\lambda_b)^2\over \prod_a dW/d\lambda_a}
={(-1)^{k(k-1)/2}\over k!(2\pi i)^k}\oint d\lambda_a{f\cdot\prod_{a<b}
(\lambda_a-\lambda_b)^2\over \prod_a dW/d\lambda_a}. \eqn\jupper$$
The integration contour in each $\lambda_a$ integral is a circle
running counterclockwise around the origin.
A factor of $k!$ comes here because, as the map from the $\lambda_a$
to the $x_i$ is of degree $k!$, each critical point of $W(x_i)$
corresponds to $k!$ critical points of $W(\lambda_a)$.

To verify that $J(f)$ is correctly normalized to coincide with $I(f)$,
we set $f=c_k(E^*)^{N-k}=\prod_a\lambda_a^{N-k}$.  According to
\conc, $I(f)=1$.  To verify that $J(f)=1$, we use the contour
integral version of \jupper.  In the denominator we can replace
$dW/\d\lambda_a=\lambda_a{}^N+(-1)^k$ by $\lambda_a{}^N-\lambda_a{}^{N-k}$
without changing the behavior on a large contour enough to affect the
integral.  Then
$$J(f)={(-1)^{k(k-1)/2}\over k!(2\pi i)^k}\oint d\lambda_1\dots
d\lambda_k{\prod_{a<b}(\lambda_a-\lambda_b)^2\over
\prod_c(\lambda_c^k-1)}.   \eqn\bulb$$
The integral is easily done as a sum of residues.  The poles
are at $\lambda_a{}^k=1$, for $1\leq a\leq k$.  Because of the
Vandermonde determinant in the numerator, the $\lambda_a$ must
be distinct.  Up to a permutation, one must have $\lambda_a=\exp(2\pi i a/k)$;
evaluating the residue at this value of the $\lambda_a$ and including
a factor of $k!$ from the sum over permutations, one gets $J(f)=1$.

\section{Quantum Field Theory Interpretation}

Physicists would never actually begin with the definition that I have
given above for the quantum cubic form.
Rather, everything begins with considerations on the
function space ${\cal W}={\rm Maps}(\Sigma,G(k,N))$.   Physicists
are mainly interested in quantum field theory, which is conveniently
formulated in terms of integration over spaces such as ${\cal W}$.

For instance, let $\Sigma$ be a complex Riemann surface with Hodge duality
operator $*$, pick a Hermitian
metric on $G(k,N)$ (such as the natural $U(k)$-invariant metric),
and for a map $\Phi:\Sigma\to G(k,N)$, set
$$L(\Phi)=\int_\Sigma(\d\Phi,*\d\Phi). \eqn\hudoxx$$
Then in the ``bosonic sigma model with target space $G(k,N)$'' we consider
integrals such as
$$\int_{{\cal W}}D\Phi\,\,\exp\left(-{L(\Phi)\over\lambda}\right),
\eqn\burdoo$$
with $\lambda$ a positive real number.  This is not complete pie in the sky.
For instance, to make the definition more concrete,
one can triangulate $\Sigma$ and make a finite dimensional approximation
to the integral.  Then the problem is to adjust $\lambda$, while
refining the triangulation, so that the given integral (and related
ones) converges as the triangulation is infinitely refined.

For a homogeneous space of positive curvature such as the Grassmannian,
one knows at a physical level of rigor precisely how to do this:
$\lambda$ must be taken to vanish in inverse proportion to the logarithm
of the number of vertices in the triangulation.  This is a consequence
of a phenomenon known as ``asymptotic freedom,'' which plays a crucial
role in the theory of the strong interactions in four
dimensions; sigma models with targets such as the Grassmannian were
intensively studied in the late 1970's and early 1980's as simple
cases of asymptotically free quantum field theories.  Asymptotic
freedom actually plays an important role in our story, since it leads
to the mass gap that will be essential in \S4.

\subsection{Supersymmetric Sigma Models}

What we actually want to do is to transfer the integral over
the space of holomorphic maps that defined the quantum cohomology
ring,
$$\langle \alpha,\beta,\gamma\rangle=\sum_\lambda\int_{{\cal M}_\lambda}
\widehat\alpha(P)\cup \widehat\beta(Q)\cup \widehat\gamma(R),\eqn\utangle$$
to an integral over the space ${\cal W}$ of all maps of $\Sigma$ to
$G(k,N)$.
Reversing the usual logic, this is done as follows.  The condition that
$\Phi:\Sigma\to G(k,N)$ is holomorphic is an equation
$$0 = \left(\bar\partial\Phi\right)^{1,0} \eqn\muxxo$$
which asserts the vanishing of a section
$$s:\Phi\to (\bar\partial \Phi)^{1,0} \eqn\uxxo$$
of an infinite dimensional vector bundle $Y$ over ${\cal W}$.
($Y$ is the bundle whose fiber at $\Phi\in {\cal W}$ is the space
of $(0,1)$ forms on $\Sigma$ with values in $\Phi^*(T^{1,0}G(k,N))$,
with $T^{1,0}G(k,N)$ being the $(1,0)$ part of the complexified
tangent bundle of $G(k,N)$.  The point of the definition is just
that $(\bar\partial\phi)^{1,0}$ is a vector in $Y$.)

\REF\mq{V. Mathai and D. Quillen, ``Superconnections, Thom Classes,
and Equivariant Differential Forms,'' Topology {\bf 25} (1986) 85.}
\def\w{\widehat}
The space ${\cal M}$ of holomorphic maps, being defined by the vanishing
of a section $s:{\cal W}\to Y$, is Poincar\'e dual to the Euler class
$\chi(Y)$ of $Y$.  So formally we can write
$$\langle\alpha,\beta,\gamma\rangle =\int_{\cal M}\w\alpha\cup\w\beta
\cup\w\gamma=\int_{\cal W}\w\alpha\cup\w\beta\cup\w\gamma\cup\chi(Y).
\eqn\mcmc$$
Now, there are any number of ways to write a differential form representing
$\chi(Y)$, but one nice way (formulated mathematically by
Mathai and Quillen [\mq])
uses a section $s$ and has a nice exponential factor $\exp(-|s|^2/\lambda)$,
with $|s|^2$ the norm of $s$ with respect to a metric on $Y$, and
$\lambda$ a positive real number.
For the section indicated in \uxxo, the norm with respect to the natural
metric is
$$|s|^2=\int_\Sigma(\d\Phi,*\d\Phi), \eqn\jxno$$
which is precisely the Lagrangian introduced above for the bosonic
sigma model with target space the Grassmannian.

\REF\topsig{E. Witten, ``Topological Sigma Models,'' Commun. Math. Phys.
{\bf 118} (1988) 411.}
\REF\atj{M. F. Atiyah and L. Jeffrey, ``Topological Lagrangians
And Cohomology,'' J. Geom. Phys. {\bf 7} (1990) 119.}
So the long and short of it is that we get a representation
$$\langle\alpha,\beta,\gamma\rangle=
\int_{{\cal W}\times \dots}\int {D}\Phi \,\,\dots \,\,
\exp\left(-{1\over \lambda}\int_\Sigma|\d\Phi|^2+\dots\right)
\w\alpha(P)\w\beta(Q)\w\gamma(R),\eqn\hardo$$
much like the bosonic sigma model, but with ``fermions,'' represented
by ``$\dots$''  The quantum field theory that appears here
is in fact a twisted form of the usual supersymmetric
nonlinear sigma model, as I explained in [\topsig]; in the twisted
model, the fermions can be interpreted in terms of differential forms
on the function space ${\cal W}$.  The relation to the Mathai-Quillen
formula was explained by Atiyah and Jeffrey [\atj] in the analogous
case of four dimensional Donaldson theory.

It should be fairly obvious that instead of $G(k,N)$ we could use
a general Kahler manifold $X$ in the above discussions, at least
at the classical level.  (If we are willing
to give up the interpretation in terms of twisting of a unitary supersymmetric
model, we can even consider almost complex manifolds that are not Kahler.)
At the quantum level, the situation is more subtle.
There are two main branches in the subject.  If $c_1(X)=0$, the
supersymmetric sigma model (with a suitable choice of the Kahler metric
of $X$) is conformally invariant; such models provide classical
solutions of string theory.  On the other hand, if $c_1>0$, as in the case
of the Grassmannian, one is in a quite different world, with asymptotic
freedom and analogs of the mass generation and chiral symmetry breaking
seen in the strong interactions.

\section{Strategy}

To try to say something of substance
in this situation, we use the realization
of $G(k,N)$ as $\mu^{-1}(0)/U(k)$, where $\mu$ is the moment map from
$\C^{kN}$ to the Lie algebra of $U(k)$.  One is tempted to try to
lift a map $\Phi:\Sigma\to G(k,N)$ to a map $\w\Phi:\Sigma\to \C^{kN}$.
There is not a natural way to do this, and there may even be a topological
obstruction.

So instead we proceed as follows.  Let $P$ be a principal $U(k)$ bundle
over $\Sigma$, $A$ a connection on $P$, $\w\Phi$ a section of
$P\times_{U(k)}\C^{kN}$, and $S$ a two-form on $\Sigma$ with values
in the adjoint bundle
${\rm ad}(P)$.  Take
$$\w
L(\w\Phi,A,S)=\int_\Sigma\left((\d_A\w\Phi,*\d_A\w\Phi)+i(S,\mu\circ\w\Phi)
\right) . \eqn\muggo$$
Then classically the theory described by $\w L(\w\Phi,A,S)$
is equivalent to the bosonic sigma model with target $G(k,N)$.
This can be seen as follows.  The Euler-Lagrange equation of $S$ is
$\mu\circ\w\Phi=0$, so, under the natural projection
$P\times_{U(k)}\C^{kN}\to \C^{kN}/U(k)$,
$\w\Phi$ maps to $\Phi:\Sigma\to \mu^{-1}(0)/U(k)=G(k,N)$.  The
Euler-Lagrange equation for $A$ identifies $P$ and $A$ with the pull-back
by $\w\Phi$ of the tautological principal $U(k)$ bundle and connection
over $G(k,N)$.  Once these restrictions and identifications are
made, $\w L(\w\Phi,A,S)$ reduces to the Lagrangian $L(\Phi)$ of the
bosonic sigma model of the Grassmannian.

This sort of reasoning is still valid quantum mechanically.
For instance, using
$$\int_{-\infty}^\infty {\d x\over 2\pi}e^{ixy}=\delta(y) \eqn\hx$$
(and the obvious generalization of that formula to several variables)
we get the path integral formula
$$\int \D S\,\,\exp\left(-i\int_\Sigma(S,\mu\circ\w\Phi)\right)
=\delta(\mu\circ\w\Phi). \eqn\hxo$$
So the $S$ integral places on $\w\Phi$ precisely the restriction
that one would guess from the classical Euler-Lagrange equations.
{}From simple properties of Gaussian integrals (which are introduced
below), one similarly deduces that, quantum mechanically as classically,
the $A$ integral has the effect of identifying $P,A$ with the pull-backs
of the tautological objects over the Grassmannian.

Similar reasoning holds after including fermions, so we get for the quantum
cubic form a representation of the general kind
$$\langle\alpha,\beta,\gamma\rangle=
\int \D \w\Phi \,\D A\, \D S\,\dots \exp\left(-\int_\Sigma\left(
(\d_A\w\Phi,*\d_A\w\Phi)+i(S,\mu\circ\w\Phi)+\dots\right)\right)
\cdot \w\alpha(P)\w\beta(Q)\w\gamma(R). \eqn\hoco$$
As before, ``$\dots$'' represents terms involving fermions that
are not indicated explicitly.

\section{Reversing The Order Of Integration}

In sum, \hoco\ will reduce to \hardo\ if we integrate over $A$ and $S$
first.  To get something interesting, we instead integrate first over
$\w\Phi$.  The key point is that $\w\Phi$ is a section of a bundle
over $\Sigma$ with linear fibers (a $\C^{kN}$ bundle)
and that $\w L$ is quadratic in $ \w\Phi$.  Consequently, the $\w\Phi$
integral is a Gaussian integral.

The basic one dimensional formula
$$\int_{-\infty}^\infty {\d x\over\sqrt {2\pi}}\exp(-\lambda x^2/2)=
{1\over\sqrt \lambda} \eqn\ncno$$
has the $n$ dimensional generalization
$$\int_{-\infty}^\infty{\d x_1\dots \d x_n\over (2\pi)^{n/2}}
\exp\left(-{1\over 2}\sum_{i,j}M_{ij}x_ix_j\right) ={1\over \sqrt {\det M}},
 \eqn\boco$$
for any quadratic form $M$ with positive real part; this is demonstrated
by picking a coordinate system in which $M={\rm diag}(m_1,\dots,m_n)$.

In our case the $\w\Phi$ integral is (apart from terms involving fermions)
$$\int\D\Phi\,\,\, \exp\left(-\int_\Sigma\left((\d_A\w\Phi,*\d_A\w\Phi)
+i(S,\mu\circ\w\Phi)\right)\right).  \eqn\gau$$
This is an infinite dimensional Gaussian integral with $M$ the quadratic
form associated with the elliptic differential operator
$$M'=(\d_A^*\d_A+iS)\otimes 1_N. \eqn\roppo$$
The notation reflects the fact that the $\C^{kN}$ bundle of which
$\w\Phi$ is a section is actually a sum of $N$ copies of a $\C^k$ bundle,
and $M$ is the sum of $N$ copies of a quadratic form derived from
an operator (namely $\d_A^*\d_A+iS$) on sections of that $\C^k$ bundle.
So the integral over $\w\Phi$
gives
$${1\over \sqrt {\det M'}}=
\left(1\over \sqrt {\det (\d_A^*\d_A+iS)}\right)^{N/2}
 .\eqn\boppo$$
The determinant of the elliptic differential operator $\d_A^*\d_A+iS$ can
be conveniently
defined using the $\zeta$-function regularization of Ray and Singer.

So modulo fermions we get
$$\langle\alpha ,\,\beta,\,\gamma\rangle=
\int\D A\,\D S\dots
\left(1\over \sqrt {\det (\d_A^*\d_A+iS)}\right)^{N/2}
 \cdot\w\alpha\w\beta\w\gamma.  \eqn\oppo$$
So we have transformed the problem of computing the quantum cohomology
of the Grassmannian to a problem involving integration over the connection
$A$ and over $S$ -- a problem in quantum gauge theory.  This brings
us into an entirely different world, that of \S2 of this paper.

As this stage we can see why -- as topologists might expect -- the sigma model
of $G(k,N)$ simplifies in the limit of $k$ fixed, $N\to\infty$.  The integrand
in \oppo\ has a sharp peak at the minimum of the determinant, and
``everything'' can be calculated in an asymptotic expansion in powers
of $1/N$, by expanding around this peak.

For fixed $N$, it is not true that ``everything'' can be calculated,
but the topological quantities can be, reducing to a saddle point
by a more elaborate argument.  The essence of the matter is that
although the classical Lagrangian $\w L$ is conformally invariant,
the quantum theory is not (because, for instance, with Ray-Singer
or any other regularization, the determinant introduced above is not
conformally invariant).  The topological quantities are however
not just conformally invariant but completely independent of the metric
of $\Sigma$.  Scaling up the metric of $\Sigma$ by a very large real factor,
life simplifies because of the basic physical properties of the model
-- asymptotic freedom and the dynamically generated mass gap.  At very
large distances (that is, if the metric on $\Sigma$ is scaled up by
a very big factor), the complicated integral over $A, S, $ and fermions
in \oppo\ reduces to a local and tractable quantum field theory
-- in fact it reduces to the gauged WZW model (of $U(k)/U(k)$)
that was analyzed in \S2.

There is a basic principle here: every quantum field theory with
a mass gap reduces at very big distances to a topological field
theory.  Often the topological field theory that so arises is more
or less trivial, but in the case of the supersymmetric sigma model
of the Grassmannian, it is the gauged WZW model.  This large distance
reduction of the Grassmannian sigma model to a gauged WZW model,
plus the relation explained in \S2 between the gauged WZW model
and the Verlinde algebra, give the relation between the quantum
cohomology of the Grassmannian and the Verlinde algebra.

It is well known that at large distances,
massive particles can be neglected and massless particles dominate.
Less fully appreciated is that beyond the reach of the propagating
fields, a non-trivial dynamics of the vacuum or topological
field theory may prevail.

\subsection{Differential Geometry Of The Moduli Space Of Bundles}

A detailed
discussion of the reduction of the sigma model to the gauged WZW model
will be the subject of \S4, but here I will make a few naive remarks.
The integrand  in \oppo\ actually has its maximum for flat connections
-- with some branching at the points $P,Q,R\in \Sigma$ at with
$\w\alpha,\w\beta, $ and $\w\gamma$ are inserted.  The moduli
space of such flat connections is (by a theorem of Mehta and
Seshadri [\seshadri]) the same as the moduli space ${\cal R}$ of
rank $k$ stable holomorphic vector bundles over $\Sigma$, with some
parabolic structure at $P,Q,R$ determined by the branching.
So the integral gives some differential geometry of ${\cal R}$.
(In view of \czonzo,
${\cal R}$ can be interpreted as a space of classical solutions
of the gauged WZW model.)  In
the large $N$ limit, direct analysis of the determinant in \oppo\ shows that
the differential geometric quantity that appears  is
the volume of the symplectic
manifold ${\cal R}$, times $N^{\dim_{{\bf C}}{\cal R}}$.
This is the leading large $N$ behavior of the
Riemann-Roch formula for the dimension of the space
$H^0({\cal R},{\cal L}^{N-k})$ of non-abelian theta functions.
This simple direct argument relates
the quantum cohomology of $G(k,N)$ to the dimension
of the space of non-abelian theta functions for large $N$.

The only way I know to establish this as an exact relation, not just as
asymptotic one for large $N$, is to reduce the sigma model of $G(k,N)$
to the gauged WZW model as we will do in \S4, and then study that model
as in \S2.

\chapter{From The Grassmannian To The $G/G$ Model}

\def\pzero{{\partial\over\partial x^0}}
\def\pone{{\partial\over\partial x^1}}
\def\cald{{\cal D}}
\REF\bagger{J. Wess and J. Bagger, {\it Supersymmetry And Supergravity},
second edition (Princeton  University Press, Princeton, 1992).}
\REF\phases{E. Witten, ``Phases Of $\N=2$ Models In Two Dimensions,''
Nucl. Phys. {\bf B403} (1993) 159 .}
This section is organized as follows.
After recalling some background about $\N=2$ models in two dimensions in \S4.1,
we construct in \S4.2 the
sigma model whose target space is the Grassmannian $G(k,N)$.
Then we analyze its behavior at long distances in \S4.3-6.  In \S4.7-8, we
enter
the computational stage and work things out in detail in the simplest
non-trivial case.

In the past, the long distance behavior of the Grassmannian sigma model
has been analyzed on ${\bf R}^2$
[\div--\cecotti];
the main results were spontaneously broken chiral
symmetry, the existence of a mass gap, and a determination for large $N$ of the
spectrum of low-lying states.  The novelty here is to examine
the long distance behavior more globally, uncovering the relation
to the gauged WZW model and thereby (in view of \S2) the Verlinde algebra.

I will make a small change in notation in this section.  In \S2,
we considered general compact Lie groups, and (as is conventional
in mathematics) we took the Lie algebra to consist of anti-hermitian
matrices (so the quadratic form $(a,b)=\Tr ab$ is negative definite).
The reason that this convention is standard for general Lie groups is
that in the case of a real group, whose representations may also all be
real, it is unnatural to introduce factors
of $i$ and therefore the group generators are naturally anti-hermitian.
In this section, the gauge group will be the unitary group
$U(k)$, which will arise in a natural complex representation,
and I will follow the standard physics convention that the
group generators are hermitian matrices; thus $(a,b)=\Tr ab$ will be positive
definite.  The complexification of the Lie algebra of $U(k)$ consists
of all $k\times k$ complex matrices; if $\sigma$ is such a matrix,
then $\bar\sigma$ will denote its hermitian adjoint.

\section{Background}

We will work in $\N=2$ superspace in two dimensions, conventions
and the basic setup
being as explained in [\bagger,\phases].
The detailed formulas of this subsection are presented mainly for reference,
and most readers will want to skim them.

We consider first flat
superspace with bosonic coordinates $x^m,\,m=0,1$ (and Lorentz
signature $-+$)
and fermionic coordinates $\theta^\alpha,\,\bar\theta
^{\alpha}$.  In a light-cone basis,
supersymmetry is realized geometrically by the
operators
$$\eqalign{ Q_{\pm} & = {\partial\over\partial\theta^\pm}+i\bar\theta^\pm
\left(\pzero\pm \pone\right) \cr
           \bar Q_\pm & = -{\partial\over\partial\bar\theta^\pm}
              -i\theta^\pm\left(\pzero\pm \pone\right).\cr} \eqn\ixox$$
These operators commute with the superspace covariant derivatives
$$\eqalign{ D_{\pm} & = {\partial\over\partial\theta^\pm}-i\bar\theta^\pm
\left(\pzero\pm \pone\right) \cr
           \bar D_\pm & = -{\partial\over\partial\bar\theta^\pm}
              +i\theta^\pm\left(\pzero\pm \pone\right)\cr} \eqn\pixox$$
which are used in constructing Lagrangians.

To formulate gauge theory, one introduces a gauge field in superspace,
replacing
the differential operators $D_\alpha$, $\bar D_\alpha$, and $\partial_m
=\partial/\partial x^m$ by gauge covariant derivatives
$\cald_\alpha$, $\bar \cald_\alpha$, and $\cald_m$.
On the superspace gauge fields one imposes the very strong constraints
$$\eqalign{0=\{\bar\cald_\alpha,\bar\cald_\beta\}&=\{\cald_\alpha,\cald_\beta\}
 \cr  \{\cald_\pm,\bar\cald_\pm\} & = 2i\left(\cald_0\pm\cald_1\right).\cr}
 \eqn\mixo$$
Among other things, these conditions permit the existence of ``chiral
superfields,'' superspace fields $\Phi$ obeying
$$ \bar\cald_\alpha \Phi = 0 .          \eqn\bixo$$

With the aid of the constraints one can take
locally
$$\eqalign{\cald_\alpha & = e^{-V}D_\alpha e^V \cr
           \bar\cald_\alpha & = e^V\bar D_\alpha e^{-V}\cr} \eqn\yixo$$
where $V$ is a real Lie algebra-valued function on superspace, called
a vector superfield.  After also fixing some residual gauge invariance,
one can go to a ``Wess-Zumino gauge,'' in which
$$\eqalign{ V & = \theta^-\bar\theta^-(v_0-v_1)+\theta^+\bar\theta^+
(v_0+v_1)-\sqrt 2\sigma \theta^-\bar\theta^+-\sqrt 2\bar\sigma\theta^+
\bar\theta^-\cr &+2i\theta^-\theta^+\left(\bar\theta^-\bar\lambda_-+\bar
\theta^+\bar\lambda_+\right)+2i\bar\theta^+\bar\theta^-(\theta^+\lambda_+
+\theta^-\lambda_-)+2\theta^-\theta^+\bar\theta^+\bar\theta^-D.\cr}
\eqn\pimxo$$
Here $v_m$ is an ordinary two-dimensional gauge field, and the other
fields are bose and fermi matter fields.  $\sigma$ is a complex $k\times k$
matrix, and -- as $V$ is hermitian -- $\bar sigma$ is its hermitian adjoint.
We write $F= F_{01}=\partial_0v_1-\partial_1v_0+[v_0,v_1]$ for
the curvature of $v$.
The supersymmetry transformation laws for this multiplet are
$$\eqalign{
\delta v_m & =i\bar\epsilon \sigma_m\lambda +i\epsilon \sigma_m\bar\lambda
                        \cr
 \delta\sigma & = -i\sqrt 2\bar\epsilon_+\lambda_--i\sqrt 2\epsilon_-\bar
             \lambda_+ \cr
\delta\bar \sigma & = -i\sqrt 2\epsilon_+\bar\lambda_--i\sqrt 2\bar\epsilon_-
             \lambda_+ \cr
\delta D & = -\bar\epsilon_+(D_0-D_1)\lambda_+-\bar\epsilon_-(D_0+D_1)\lambda_-
 + \epsilon_+(D_0-D_1)\bar\lambda_++\epsilon_-(D_0+D_1)\bar\lambda_-
\cr &+\sqrt 2 \epsilon_+[\sigma,\bar\lambda_-] +\sqrt 2\epsilon_-[\bar\sigma,
\bar\lambda_+]+\sqrt 2[\sigma,\lambda_+]\bar\epsilon_-+\sqrt 2
[\bar\sigma,\lambda_-]\bar\epsilon_+ \cr
\delta\lambda_+ & = i\epsilon_+D+\sqrt 2(D_0+D_1)\bar\sigma\epsilon_-
         -F_{01}\epsilon_+ -[\sigma,\bar\sigma]\epsilon_+ \cr
\delta\lambda_- & = i\epsilon_-D+\sqrt 2(D_0-D_1)\sigma\epsilon_+
         +F_{01}\epsilon_- +[\sigma,\bar\sigma]\epsilon_- \cr
\delta\bar\lambda_+ & = -i\bar\epsilon_+D+\sqrt 2(D_0+D_1)\sigma\bar
\epsilon_-  -F_{01}\bar\epsilon_+ +[\sigma,\bar\sigma]\bar\epsilon_+ \cr
\delta\bar\lambda_- & = -i\bar\epsilon_-D+\sqrt 2(D_0-D_1)\bar\sigma
\bar\epsilon_++F_{01}\bar\epsilon_- -[\sigma,\bar\sigma]\bar\epsilon_-. \cr}
         \eqn\milopo$$

The basic gauge invariant field strength is
$$\eqalign{\Sigma ={1\over 2\sqrt 2}\{\bar \cald_+,\cald_-\}   &
 = \sigma+i\sqrt 2\theta^+\bar\lambda_+-i\sqrt 2\bar\theta^-\lambda_-
+\sqrt 2\theta^+\bar\theta^-D\cr &-i\bar\theta^-\theta^-(D_0-D_1)\sigma
-i\theta^+\bar\theta^+(D_0+D_1)\sigma \cr &
+\sqrt 2\bar\theta^-\theta^-\theta^+
(D_0-D_1)\bar\lambda_+-\sqrt 2\theta^+\bar\theta^+\bar\theta^-
(D_0+D_1)\lambda_--i\sqrt 2\theta^+\bar\theta^- F_{01}
\cr & -2i\theta^-\bar\theta^-\theta^+[\sigma,\bar\lambda_-]
-2i\bar\theta^-\theta^+\bar\theta^+[\sigma,\lambda_+]\cr &
-\bar\theta^-\theta^-\theta^+\bar\theta^+\left((D_0{}^2-D_1{}^2)\sigma
-[\sigma,[\sigma,\bar\sigma]]\right)
+i\theta^-\bar\theta^-\theta^+\bar\theta^+[\sigma,\partial_mv^m]
.\cr} \eqn\nogof$$
(The last term does not really spoil gauge invariance:
the gauge transformations that preserve Wess-Zumino
gauge have a certain $\theta$ dependence which requires this term to be
present.)
$\Sigma$ is a twisted chiral superfield; this means that
(by the Bianchi identity together with the constraints)
$$\bar\cald_+\Sigma=\cald_-\Sigma = 0 . \eqn\hutcho$$

With the aid of $\Sigma$,
it is straightforward to construct gauge invariant
Lagrangians.  The standard gauge kinetic energy is
$$\eqalign{L_g & =-{1\over 4e^2}\int d^2x\,\,\d^4\theta \Tr\bar\Sigma\Sigma
 \cr &
={1\over e^2}\int\d^2x\,\Tr
\left({1\over 2}F_{01}^2+|D_0\sigma|^2-|D_1\sigma|^2
+i\bar\lambda_-(D_0+D_1)\lambda_-+i\bar\lambda_+(D_0-D_1)\lambda_+
\right.\cr &~~~~~~~\left.+{1\over 2}D^2-{1\over 2}[\sigma,\bar\sigma]^2
-{\sqrt 2}\lambda_+
[\sigma,\bar\lambda_-]
+{\sqrt 2}[\bar\sigma,\lambda_-]\bar\lambda_+\right)
.\cr}\eqn\muccdo$$

One more term constructed from gauge fields only is important.
Using the fact that $\Sigma$ is a twisted chiral superfield, there
is an invariant interaction of the form
$$L_{D,\theta}={it\over 2\sqrt 2}\int\d^2 x \,\d\theta^+\,d\bar\theta^-
\,\,\Tr \Sigma|_{\theta^-=\bar\theta^+=0}+{\mit c.c.}
=\int\d^2 x \left(-r\Tr D+{\theta\over 2\pi}\Tr F_{01}\right), \eqn\jumbox$$
with
$$t=ir+{\theta\over 2\pi}.        \eqn\uccdo$$

\subsection{Matter Fields}

Chiral superfields are functions $\Phi$ on superspace,
transforming in some given unitary representation $V$ of the gauge group,
and obeying
$$\bar\cald_\pm \Phi = 0 .        \eqn\ruccdo$$
Such a field has an expansion in components
$$\Phi = \phi +\sqrt 2\theta^\alpha\psi_\alpha +
    \theta^\alpha\theta_\alpha F        .      \eqn\wuddo$$

The supersymmetry transformation laws for this multiplet are (by
dimensional reduction from [\bagger, p. 50])
$$\eqalign{
\delta \phi & = \sqrt 2\left(\epsilon_+\psi_--\epsilon_-\psi_+\right) \cr
\delta \psi_+ &= i\sqrt 2\left(D_0+D_1\right)\phi\bar\epsilon_-
+\sqrt 2\epsilon_+ F - 2 \bar\sigma\phi\bar\epsilon_+ \cr
\delta \psi_- &= -i\sqrt 2\left(D_0-D_1\right)\phi\bar\epsilon_+
+\sqrt 2\epsilon_- F + 2 \sigma\phi\bar\epsilon_- \cr
\delta F & = -i\sqrt 2\bar\epsilon_+\left(D_0-D_1\right)\psi_+
-i\sqrt 2\bar\epsilon_-\left(D_0+D_1\right)\psi_-
\cr & ~~~+2\left(\bar\epsilon_+\bar\sigma\psi_-+\bar\epsilon_-\sigma\psi_+
\right) + 2i\left(\bar\epsilon_-\bar\lambda_+-\bar\epsilon_+\bar
\lambda_-\right)\phi
. \cr}
\eqn\nurbob$$

The usual kinetic energy for a multiplet of such chiral superfields is
$$\eqalign{L_{{\mit ch}}={1\over 4}\int d^2x \,d^4\theta\,\,\bar\Phi \Phi
= & \int d^2x\,\,\left(|D_0\phi|^2-|D_1\phi|^2 +|F|^2
+i\bar\psi_+(D_0-D_1)\psi_+\right.\cr &\left.+i\bar\psi_-(D_0+D_1)\psi_-
+\bar\phi D\phi -\bar\phi\{\sigma,\bar\sigma\}\phi
\right.\cr & \left.-\sqrt 2\bar\psi_+\bar\sigma\psi_-
-\sqrt 2\bar\psi_-\sigma\psi_+
+i\sqrt 2 \bar\psi_+\bar\lambda_-\phi -i\sqrt 2 \bar\psi_-\bar\lambda_+\phi
\right.\cr & \left.
+i\sqrt 2\bar\phi \lambda_+\psi_- -i\sqrt 2\bar\phi\lambda_-\psi_+\right).
\cr}  \eqn\qqmmw$$

\REF\hitchin{N. J. Hitchin, A. Karlhede, U. Lindstrom, and M. Rocek,
``Hyperkahler Metrics And Supersymmetry,'' Commun. Math. Phys. {\bf 108}
(1987) 535.}
The $|D_\alpha \phi|^2$ term in \qqmmw\ is the conventional
free kinetic energy corresponding to a sigma model with a flat
metric on $V\cong {\bf C}^r$.
The $\bar\phi D\phi$ term is the coupling of $D$ to the moment map,
in the sense that if we pick a basis $T_a,\, a=1
\dots \dim G$ for the Lie algebra of $G$, then this term
is
$$\int \d^2x \sum_a D^a (\bar\phi,T_a\phi), \eqn\iglo$$
and the functions $(\bar\phi,T_a\phi)$ are the components of the
moment map.

A more general Kahler metric on $V$ (and
accordingly, a more general form of the moment map)
could be obtained by replacing the function $\bar\Phi\Phi$ on the left
hand side of \qqmmw\ with a more general Kahler potential $K(\Phi,\bar\Phi)$.
These matters are explained in some detail in [\hitchin].

\section{The Model}

Now we can construct the actual model of interest.  We take the
gauge group to be $G=U(k)$.  We take $kN$ chiral superfields
$\Phi^{is}$, $i=1\dots k$, $s=1\dots N$, regarded as $N$ copies
of the defining $k$ dimensional representation of $G$.
The action of $G$ commutes with a global
symmetry group $H\cong U(N)$ which one can think of as the unitary
transformations of ${\bf C}^N$.

The Lagrangian that we actually wish to study is simply
$$L = L_{{\mit gauge}}+L_{D,\theta}+L_{{\mit ch}}. \eqn\ripporo$$
The potential energy is determined
by the following terms in $L$:
$$L_{{\mit pot}} ={1\over 2e^2}\Tr D^2-r\Tr D + \bar\phi D\phi
-{1\over 2e^2}\Tr[\sigma,\bar\sigma]^2- \bar\phi\{\sigma,\bar\sigma\}\phi.
       \eqn\pixox$$
Upon integrating out $D$, the
potential energy is
$$V = {e^2\over 2}\sum_{i,j=1}^k\left(\sum_s\bar\phi_{is}\phi^{js}-\delta_i{}^j
r\right)^2+{1\over 2e^2}\Tr[\sigma,\bar\sigma]^2
+\bar\phi\{\sigma,\bar\sigma\}\phi. \eqn\jixox$$

The space of classical vacua is the space of zeroes of $V$ up to gauge
transformation.  For $V$ to vanish, $\phi$ must be non-zero, and
therefore $\sigma$ must vanish.    As anticipated in \S3, the first
term in the potential is the square of the moment map for the action
of $U(k)$ on ${\bf C}^{kN}$.
This term vanishes precisely if the vectors in ${\bf C}^N$ represented
by the rows of $\phi$, divided by $\sqrt r$,  are orthonormal.
The $k$ dimensional subspace $V\subset {\bf C}^N$ spanned by the rows of $\phi$
is gauge invariant, and is the only gauge invariant data determined
by $\phi$ (since any two orthonormal bases of $V$ are related
by the action of $U(k)$).  Moreover, every $k$ dimensional subspace
$V\subset {\bf C}^N$   has such an orthonormal basis.
So the space of classical vacua is the Grassmannian $G(k,N)$ of
$k$ dimensional subspaces of ${\bf C}^N$.

Since the condition for vanishing energy is
$$\sum_s\bar\phi_{is}\phi^{js}=\delta_i{}^jr, \eqn\kop$$
the radius of the space of vacua is $\sqrt r$ and the Kahler
class is proportional to $r$.
Classically, the space of vacua shrinks to a point for $r=0$; for
$r<0$ the classical energy can no longer vanish and it appears
that supersymmetry is spontaneously broken.  Quantum
mechanically, the situation is rather different and there is a smooth
continuation to negative $r$ with unbroken supersymmetry,
as discussed (for $k=1$) in [\phases,\S3.2];
the existence of this continuation will be exploited below.

The choice of a classical vacuum spontaneously breaks the symmetry
group $U(N)$ to $U(k)\times U(N-k)$, while leaving supersymmetry
unbroken.  The oscillations in the vacuum
are massless Goldstone bosons at the classical level.  Their
supersymmetric partners are, of course, also massless classically.
Other modes are readily seen to have masses proportional to $e$.
The model therefore reduces at long distances (or equivalently
for $ e\to\infty$) to the supersymmetric nonlinear sigma model
with target space the Grassmannian; we will more briefly call this
the Grassmannian sigma model.

At the quantum level, spontaneous breaking of a continuous
symmetry such as the $U(N)$ symmetry of this model is not possible
in two dimensions.  The symmetry must therefore be restored by
quantum corrections.  Exhibiting this symmetry restoration, and the
associated mass gap, was a primary goal of early investigations
of the model.

\subsection{$R$ Symmetries}

A right-moving $R$-symmetry in an $\N=2$ model in two dimensions is
a symmetry under which $\theta^+\to e^{i\alpha}\theta^+,$
$\bar\theta^+\to e^{-i\alpha}\bar\theta^+$, while $\theta^-,
\bar\theta^-$ are invariant.  A left-moving $R$-symmetry
obeys the analogous condition with $\theta^+$ and $\theta^-$
exchanged.

The Grassmannian sigma model as constructed above has at the classical
level a right-moving
$R$-symmetry $J_R$\foot{We will somewhat imprecisely use the symbol
$J_R$ to denote either the current or the corresponding charge;
and similarly for other currents introduced momentarily.}
under which the charges of the various fields
are as follows: $(\psi_+,F,\sigma,\lambda_-)$ have charges
$(-1,-1,1,1)$, their complex conjugates have opposite charge,
and other fields have charge zero.  Similarly there is classically
a left-moving $R$-symmetry $J_L$ under which $(\psi_-,F_i,\sigma,\lambda_+)$
have charges $(-1,-1,-1,1)$, their complex conjugates have opposite
charges, and other fields are neutral.

At the quantum level, the sum $J_V=J_R+J_L$ is a ``vector'' symmetry,
that is, it transforms left- and right-moving fermions the same way,
so it is free of anomaly and generates a $U(1)$ symmetry.

However, the ``axial'' combination
$J_A=J_R-J_L$ is anomalous.
The anomaly can be described as follows.
Let $a$ be a $U(1)$ connection with first Chern class 1,
and embed this in $G=U(K)$ so that the $U(k)$ gauge
field is $v={\rm diag}(a,0,0,\dots,0)$.  In such an instanton
field, the index of the Dirac operator acting on $\psi_+$ is $N$.
\foot{The index is defined as
the number of $\psi_+$ zero modes minus the number of
$\bar\psi_+$ zero modes. Recall that $\psi_+$ transforms
as a sum of $N$ copies of the defining $k$-dimensional representation
of $U(k)$; each of these contributes 1 to the index.}
Similarly the $\psi_-$ index is $-N$.  The total anomaly in
$J_A=J_R-J_L$ is the difference of these or $2N$.
The anomaly in any instanton field would be an integer multiple of this.

So  $J_A$ is conserved only modulo $2N$.
The only symmetries we can construct from $J_A$ are the
discrete transformations $\exp(2\pi it J_A/2N)$, with $t\in {\bf Z}$.
This gives a discrete group, isomorphic to ${\bf Z}_{2N}$,
of chiral symmetries.  If unbroken, these symmetries would
prevent $\psi$ and $\lambda$ from gaining a mass.  One of the
main results of the old literature on this model was that
this ${\bf Z}_{2N}$ is spontaneously broken down to ${\bf Z}_2$,
making a mass gap possible;
the surviving ${\bf Z}_2$ is just the operation $(-1)^F$
that counts fermions modulo two.

\subsection{The Twisted Model}

\REF\witmir{E. Witten, ``Mirror Manifolds And Topological Field
Theory,'' in {\it Essays On Mirror Manifolds}, ed. S.-T. Yau
(International Press, 1992).}
Any $\N=2$ supersymmetric theory in two dimensions with an $R$ symmetry
can be twisted to obtain a topological field theory.
The construction, as explained in [\witmir], which the reader can consult
for details,  involves
adding to the usual stress tensor
the derivative of the $R$-current.
As we have just seen, in the case of the Grassmannian there is
only one anomaly-free $R$-symmetry.   Consequently,
only one twisted topological field theory can be constructed;
it is related to the quantum cohomology of the Grassmannian,
which was introduced in \S3.

In going from the untwisted to the twisted model, the spin of
every field decreases (in the convention of [\phases])
by $J_V/2$.  For instance, in the untwisted
model, $\psi_+$ and $\bar\psi_+$ have spin $1/2$ and $J_V=\mp 1$,
so in the twisted model they have respectively spin $1$ and 0.
More generally the fermi fields that have spin zero  in the twisted model
are $\bar\psi_+,\psi_-,\bar\lambda_-,\lambda_+$.
If the twisted model is formulated on ${\bf S}^2$, the spin zero fields
each have one zero mode and the spin one fields have none.
Since $\bar\psi_+,\psi_-,\bar\lambda_-,\lambda_+$ have
$kN,kN,k^2,k^2$ components respectively and have $J_A=1,1,-1,-1$,
the total $J_A$ value of the zero modes is $kN+kN-k^2-k^2=2k(N-k)$,
and this is the anomaly in $J_A$ conservation due to coupling to the
curvature of $S^2$.  (Not coincidentally, $2k(N-k)$ is the
dimension of $G(k,N)$.)  More generally, on a surface of genus $g$, the
spin one fields would have $g$ zero modes, so the violation of $J_A$ is
$$\Delta J_A = 2k(N-k)(1-g)=k(N-k)\int_\Sigma d^2x\sqrt h {R\over 2\pi}.
                    \eqn\oxoc$$
Here I have written the Euler characteristic of $\Sigma$, which of course
equals $2(1-g)$, as the familiar curvature integral.

\subsection{Fermionic Symmetry Of The Twisted Theory}

The untwisted theory, formulated on a flat world-sheet, possesses
fermionic symmetries, that were described in detail in
equations \milopo, \nurbob.  After twisting, the fermionic parameters
$\epsilon_+$ and $\bar\epsilon_-$ in the transformation laws have
spin zero; let
$Q_-$ and $\bar Q_+$ be the symmetries generated by those transformations,
and let $Q=Q_-+\bar Q_+$.
By a standard calculation, $Q_-{}^2=\bar Q_+{}^2=\{Q_-,\bar Q_+\}=0$
and in particular $Q^2=0$.
Moreover, the stress tensor
can be written as $T=\{Q,\Lambda\}$ for some $\Lambda$.
It follows that if we restrict ourselves to operators that
are annihilated by $Q$ (or more exactly to cohomology
classes of such operators), the theory can be interpreted
as a topological field theory.  Each cohomology class of $Q$-invariant
operators has representatives annihilated by both $Q_-$ and $\bar Q_+$.

The relevant observables are easily found.  The transformation laws of the
topological   theory are found from the microscopic transformation laws
\milopo, \nurbob\ by setting $\epsilon_-=\bar\epsilon_+=0$ and
keeping $\epsilon_+,\bar\epsilon_-$.  By inspection of the transformation
laws, $\sigma$ (but not $\bar\sigma$)
is invariant, so that any gauge invariant holomorphic
function of
$\sigma$ is a suitable vertex operator in the topological theory.
Such functions are linear combinations of characters, so the
basic operators constructed this way are
$$O_V(x)=\Tr_V\sigma(x) , \eqn\burfo$$
with $V$ an irreducible representation of $G=U(k)$ and $\Tr_V$ the
trace in that representation.

Actually, these are the only relevant operators.  In fact, even
before twisting, the model (for $r>>0$)
is equivalent at long distances, as we saw above,
to a sigma model with target the Grassmannian $G(k,N)$.
Consequently, the twisted model is simply the standard $A$
model of $G(k,N)$ (the $A$ model for any Kahler target is explained
in detail in [\witmir])
so the cohomology classes of observables are in one-to-one correspondence
with the de Rham cohomology of $G(k,N)$.

Indeed, upon integrating out the massive fields, $\sigma(x)$ turns
into a bilinear expression in massless fermions tangent to $G(k,N)$,
with values in the adjoint representation of $U(k)$.
It is easy to calculate this explicitly in the weak coupling, low energy
limit.  To this aim, we need only evaluate a tree diagram, and
we can ignore the kinetic energy of the massive $\sigma$ field.  The
relevant part of the Lagrangian is simply
$$-\int d^2x\left(\sum_{ijs}\bar\phi_{is}\{\sigma,\bar\sigma\}^i{}_j
\phi^{js} +\sqrt 2\sum_{ijs}\bar\psi_{+is}\bar\sigma^i{}_j\psi_-{}^{js}
\right).                          \eqn\purrify$$
Because of the overall $U(N)$ invariance, it suffices to work out
the effective operator representing $\sigma$ in the low energy theory
at one particular point on $G(k,N)$.  We take this to be the point
represented by $\phi^{is}=\sqrt r \delta^{is}$ for $1\leq s\leq k$,
$\phi^{is}=0$ for $s>k$.  With this choice, \purrify\ becomes
$$-\int d^2x \left(2r\Tr \bar\sigma\sigma +\sqrt 2 \sum_{ijs}
\bar\psi_{+is}\bar\sigma^i{}_j\psi_-{}^{js}\right).\eqn\durify$$
The quickest way to evaluate the tree diagram is simply to impose
the equation of motion of $\bar \sigma$; this gives
$$ \sigma^j{}_i=-{1\over r\sqrt 2}\sum_s\bar\psi_{+is}\psi_-^{js}.
\eqn\reddo$$
In the interpretation of the low energy theory in terms of differential
forms on $G(k,N)$, $\psi_-/\sqrt r$ and $\bar\psi_+/\sqrt r$
are $(1,0)$ and $(0,1)$ forms.  ($\psi_-$ and $\bar\psi_+$ have
been normalized to have canonical kinetic energies; their natural
normalization as differential forms involves dividing by $\sqrt r$.)
So $\sigma$ is represented in the low energy theory by a $(1,1)$
form or more exactly by the operator in the $G(k,N)$ model
determined by this $(1,1)$ form.

The chosen vacuum $\phi^{is}=\sqrt r\delta^{is}$ is invariant up to
a gauge transformation under a subgroup $U(k)\times U(N-k)$ of $U(N)$.
For $\sigma$ to be a $U(N)$-invariant form in the adjoint representation
of the gauge group, it must transform in the adjoint
representation of the unbroken $U(k)$ (since the unbroken symmetry
is a mixture of this with a gauge transformation)
and be invariant under $U(N-k)$.  The $U(N)$ action can then be
used to extend $\sigma$ in a unique way to an invariant $(1,1)$ form
on $G(k,N)$.  It is evident that
the right hand side of \reddo\ has the required properties.

Conversely, the right hand side of \reddo\ is the only bilinear expression
in $\bar\psi_+$ and $\psi_-$ with the claimed properties, so any
adjoint-valued $U(N)$-invariant $(1,1)$ form would be a multiple of $\sigma$.
Such a form is the curvature of the tautological $U(k)$ bundle
$E^*$ with its natural connection.
So up to a  constant, which I will not verify
directly (it can be absorbed in the constant later called $c$),
$\sigma$ coincides in the low energy theory
with the tautological curvature.  Hence classical
expressions $O_V= \Tr_V\sigma$ coincide with the corresponding polynomials
in Chern
classes on $G(k,N)$, and as quantum operators in the twisted theory,
the $O_V$ coincide with the elements of the quantum cohomology determined
by those classes.
The fact that the tautological classes generate the cohomology
of $G(k,N)$ ensures that the $O_V$ span the space of observables of
the twisted theory.

There is a more conceptual approach to identifying $\sigma$ with the
tautological curvature which I will indicate very briefly.
Let $\lambda_-=\eta_0-\eta_1$,
$\bar\lambda_+=\eta_0+\eta_1$.  Restrict to the diagonal fermionic
symmetry with $\epsilon_+=\bar\epsilon_-=\epsilon$.  Then
a key part of the symmetry algebra is
$$\eqalign{\delta v_m & = 2i\epsilon \eta_m \cr
           \delta \eta_m & =\sqrt 2\epsilon D_m\sigma \cr
           \delta \sigma & = 0. \cr}         \eqn\iffo$$
This multiplet describes the equivariant cohomology of the gauge
group acting on the space ${\cal A}$ of connections.
The interpretation of $\sigma$ as the curvature of the tautological
bundle over the quotient is standard in equivariant cohomology.
This interpretation holds independent of any specific Lagrangian model;
the salient feature of
the particular model we are considering is that the connection $v_m$
is identified via the low energy equations of motion with the pullback
of the tautological connection on $E^*\to G(k,N)$.

\subsection{Instantons}

A correlation function
$$\left\langle \prod_i O_{V_i}(x_i)\right\rangle \eqn\guelf$$
on a Riemann surface $\Sigma$ can be computed as follows.
The $O_{V_i}$ determine cohomology classes of $G(k,N)$ as we have just
seen; pick Poincar\'e dual cycles $H_i$.
Let $d$ be the non-negative integer, if any,
 such that the moduli space of holomorphic
maps $\Phi:\Sigma\to G(k,N)$ obeying
$$\Phi(x_i)\in H_i \eqn\lilko$$
has virtual dimension zero.  The correlation function \guelf\
is zero if such a $d$ does not exist; otherwise it is
$$\left\langle\prod_i O_{V_i}(x_i)\right\rangle = \exp(-dr)\cdot N_{\{H_i\}}
\eqn\mbmb$$
with $N_{\{H_i\}}$ the ``number'' of holomorphic maps $\Phi:\Sigma
\to G(k,N)$ that obey \lilko.
(In general, in defining this number,
one must make a suitable perturbation of the equation to
avoid possible degeneracies; that is why I have put the word ``number''
in quotes.)
\mbmb\ follows from the standard description of the
$A$ model, as explained in [\witmir].

More microscopically, to see the appearance of instantons,
one can begin with the transformation
laws \milopo, \nurbob.  The calculation
of the correlation function in
\guelf\ can be localized, by a standard argument, on the fixed
points of $Q_-,\bar Q_+$.  An analysis as in [\phases], pp. 184-8,
identifies those fixed points (for $r>>0$)
with the holomorphic maps of $\Sigma$ to the Grassmannian.
Those holomorphic maps appear in precisely the form in which they
were studied by Bertram, Daskapoulos, and Wentworth [\bertram].

\section{Some Renormalization Factors}

Before analyzing the quantum theory, I want to first
point out a few details involving renormalization.

Any topological field theory in two dimensions could be modified
by the addition of a term
$$\Delta L = a\int_\Sigma d^2x\sqrt h {R\over 2\pi} \eqn\normert$$
without affecting the topological invariance.  The affect of this
is merely to multiply
a genus $g$ amplitude by a factor of $\exp(a(2-2g))$.

An important role in the analysis will be played by the Kahler
parameter $r$.  For instance, $r$ enters
in the basic formula \mbmb\ expressing correlation functions in
terms of instantons.  However, as we see in \mbmb, the $r$ dependence
of a degree $d$ instanton contribution is known  {\it a priori}.
Moreover, because of the ${\bf Z}$ grading of the classical cohomology,
every given correlation function in genus $g$ receives a contribution
at most only from one known value of $d$.  Therefore, there is no material
loss in setting $r$ to 0, and we will do that eventually in \S4.7.

Another normalization question involves the possibility of multiplying
an operator of degree $w$ by a factor $\exp(uw)$ with some constant $u$.
One can show by keeping track of the classical ${\bf Z}$ grading that
this can be absorbed in adding constants to $r$ and $a$.
This normalization question will arise below because we will
find that the field $\sigma$ of the Grassmannian sigma model
has a macroscopic interpretation as
$$\sigma=c g,\eqn\juppl$$
where $c$ is a constant
that we will determine only approximately and $g$ is the elementary field
of a gauged WZW model.

In practice, in our computations we will not try to determine
the precise values of $a$ and $c$.  At the end, when we enter
the computational stage, we will identify the values of these
parameters by checking a couple of special cases of the formulas.

\section{Quantum Properties Of The Model}

We come finally to the point of the present paper -- the calculation
mapping the Grassmannian sigma model onto the $G/G$ model,
and thence the Verlinde algebra.
The calculations themselves are not new [\div--\cecotti],
and I will therefore present them rather briefly.  What is new
is the result that we will get by considering these computations
in a global context.

We begin with the expression for $D$ in terms of matter fields
that is obtained by varying the potential energy term \pixox\ with
respect to $D$:
$$-{1\over e^2}D^i{}_j=\sum_{s=1}^N\phi^{is}\bar\phi_{js}-r\delta^i_s.
       \eqn\oppu$$
At the classical level, for $r>>0$, vanishing of the $D^i{}_j$ -- which
is needed to set the energy to zero -- requires that the $\phi^{is}$
should have non-zero vacuum expectation values.  This in turn spontaneously
breaks the global $U(N)$ symmetry (and ensures the existence of
massless Goldstone bosons and the absence of a mass gap).  Such
spontaneous breaking of a continuous symmetry
is, however, impossible in two dimensions.

The resolution of this conundrum has long been known.  Quantum
mechanically the operator $O^i{}_j=\sum_{s=1}^N\phi^{is}\bar\phi_{js}$ can
have an expectation value even if the $\phi^{is}$ do not.  If
this expectation value can equal $r\delta^i{}_j$, then the $D^i{}_j$
can vanish without spontaneous breaking of the $U(N)$ symmetry.

To investigate this phenomenon, let us compute the expectation value
of $O^i{}_j$.  We will first do this in a naive approximation,
treating the $\phi$'s as free fields with the mass term
that we can read off from the classical Lagrangian.  Then we will
discuss the conditions for validity of the approximation.
We will do the calculation on Euclidean ${\bf R}^2$, making the
standard Wick rotations from the Lorentz signature Lagrangian
given above.

The mass term for the $\phi$ field in the Lagrangian
is $\sum_{i,j,s}\bar\phi_{is}\{\sigma,\bar\sigma\}^i{}_j\phi^{is}$, with
$\{\cdot,\cdot\}$ the anticommutator.
Treating the $\phi$'s as free fields with that mass term,
the expectation value of $O^i{}_j$ is simply
$$\langle O\rangle =N\int {d^2k\over (2\pi)^2}{1\over k^2+\{\sigma,\bar\sigma
\}}.\eqn\tugboat$$
The factor of $N$ comes from summing over $s$.

The integral in \tugboat\ is logarithmically divergent.  The divergence
can be regularized by subtracting a similar integral with
$\{\sigma,\bar\sigma\}$ replaced by a multiple of the identity,
say $2\mu^2$, with $\mu$ an arbitrary ``subtraction point.''
The subtraction can be interpreted as an additive renormalization of
$r$.  After this regularization, the integral can be evaluated,
and one gets
$$\langle O\rangle = -{N\over
4\pi}\ln\left(\{\sigma,\bar\sigma\}/2\mu^2\right).
 \eqn\wondrous$$
The condition for $D$ to vanish in this approximation is hence that
$$-{N\over 4\pi}\ln\left(\{\sigma,\bar\sigma\}/2\mu^2\right)-r =
0,\eqn\ugboat$$
or
$$\{\sigma,\bar\sigma\}= 2\mu^2\exp\left(-4\pi r/N\right). \eqn\marblo$$

This is however only a necessary condition for vanishing of the energy.
Another condition comes from the presence in the classical
Lagrangian of a term proportional to $\Tr[\sigma,\bar\sigma]^2$.
This term gives a contribution to the energy that
vanishes precisely when $[\sigma,\bar\sigma]=0$,
so in seeking to describe the vacuum, we may assume that $\sigma$
and $\bar\sigma$ commute and therefore rewrite \marblo\ in the form
$$\sigma\bar\sigma=\mu^2\exp\left(-4\pi r/N\right). \eqn\tarble$$
This means that
$$\sigma = c g \eqn\arble$$
with $g$ a unitary matrix -- $g = 1$ -- and
$c$ the constant
$$c =\mu\exp\left(-2\pi r/N\right). \eqn\ommo$$
Thus, we have obtained a kind of sigma model with a field
$g$ taking values in the unitary group.

The vacuum expectation value of $\sigma$ that we have just found
gives a positive mass squared to the $\phi^{is}$, so that they
will have zero vacuum expectation value, restoring the $U(N)$
symmetry.  However, the discrete chiral symmetry (conservation
of $J_A$ modulo $2N$) is spontaneously broken in this process.
Indeed, since $\sigma$ has $J_A=2$, the vacuum expectation
value of $\sigma$ breaks ${\bf Z}_{2N}$ down to ${\bf Z}_2$.
(For instance, this is discussed in detail for $k=1$ on p. 310
of [\oldwit].)
As the broken symmetry is discrete, this does not produce
Goldstone bosons and is compatible with the existence of a mass
gap.  In fact, the broken symmetry helps in getting a mass gap,
since most of the fermions obtain masses at tree level from the
vacuum expectation value of $\sigma$.

\subsection{Validity Of The Approximation}

Before proceeding to unravel further subtleties, let us
discuss the conditions for validity of the approximation.

The traditional region of validity of the above approximation,
as in [\div,\oldwit], is $k$ fixed, $N\to \infty$, with $r$ and $1/e^2$
of order $N$.  In this limit, the corrections to the
approximation (of treating the $\phi$'s as free fields with
a $\sigma$-dependent mass) are of order $1/N$.  The above
computation is part of the beginning of a systematic
expansion of all physical observables in powers of $1/N$.
Many important features of the theory
involve properties that are stable under perturbation -- like
whether there is a mass gap, what symmetries are spontaneously
broken, and certain aspects of the topological
sector.  For addressing such questions, the $1/N$ expansion is
good enough for fixed $k$ and sufficiently big $N$.

That is not enough for us, because we wish
to relate the Verlinde algebra to the cohomology
of the Grassmannian for all $k$ and $N$.  Happily, there is another
region of validity of the approximation.  At the classical level,
the matrix $O$ is positive definite, and accordingly for $r<0$
it would be impossible for the energy to vanish.  Quantum mechanically,
because of the subtraction that was needed in the above computation,
$O$ is not positive definite.  Accordingly, zero energy is possible
also for negative $r$ at the quantum level; indeed, the solution
\tarble\ makes sense for either sign of $r$.  (The continuation of the
model to negative $r$ was discussed in [\phases,\S3.2] for the case
$k=1$.)

I claim that for any $k$ and $N$, the computation leading to \tarble\
is a valid approximation for $r<<0$.  The reason for this is that
the approximate vacuum state given by this computation has
exponentially {\it large} $\sigma$ for $r\to\-\infty$.  This gives
an exponentially large mass to the $\Phi$ multiplet, so $\phi$ and $\psi$ loops
can be ignored except perhaps for renormalization effects involving
diagrams with poor ultraviolet convergence.  In this super-renormalizable
theory, the only such diagram is the one loop diagram whose evaluation
leads to \tarble.

The quantum cohomology of the Grassmannian involves, naively,
the behavior for $r>>0$.  However, because the first Chern class
of $G(k,N)$ is positive, every topological correlation function of the
twisted theory (of operators
of definite dimension or $J_A$) receives a contribution only from
one value of the instanton number and hence depends on $r$
as $\exp(-dr)$ with a known constant $d$ that appeared in \mbmb.
The behavior for $r<<0$
therefore determines the behavior for $r>>0$.  Consequently,
the fact that our approximation is valid for the theory continued
to $r<<0$ means that it is good enough for studying the topological
sector of the twisted theory.  We will now explore the
implications.

\section{The Mass Gap And The WZW Action}

Because $\sigma$ was determined to be an arbitrary unitary matrix
(times a fixed constant), it appears at first sight that
the model has a continuous vacuum degeneracy and therefore
massless particles, at least in this approximation.  This can
hardly be correct because the massless $\sigma$ particles,
subject to the constraint \tarble, do not furnish a representation
of $\N=2$ supersymmetry.  (The $\phi$ fields are massive in this
approximation,
as we have noted, and so cannot help.)  This puzzle was resolved in the
old literature in the context of the $1/N$ expansion;
the resolution involves giving a mass to $\sigma$ by mixing with the
$U(k)$ gauge field $v$.
(See, for instance, p. 308 of [\oldwit] for $k=1$ and
the discussion of the $\phi_5-\lambda$ propagator
on pp. 165-6 of [\brazil] for general $k$.)
I will not present the detailed computations here, as they are standard;
I will merely summarize them and focus on the interpretation, which is all
that is new.

\FIG\fermloop{The one-loop diagram describing $\sigma - v$ mixing}
\FIG\ofermloop{The one-loop diagram generating the Wess-Zumino coupling
for $\sigma$.}
The $1/N$ expansion amounts to integrating out
the chiral superfields $\Phi^{is}$ to obtain
an effective action for the gauge multiplet.
The $\sigma - v$ mixing comes from the one loop diagram of figure (\fermloop).
The non-vanishing contribution
is the one in which the particles running around the loop are fermions.
However, perhaps even more fundamental is the one-loop diagram with
external sigma fields only and internal fermions, shown in figure
(\ofermloop).
The fermions $\psi^{is}$, for $s=1\dots N$, form $N$ copies of the
fundamental representation of $U(k)$.  Let us suppress the $s$ index
and look at a single multiplet $\psi^i$.  The key point is that,
looking back to the Lagrangian
\qqmmw, the fermions receive their mass from a coupling $-\sqrt 2\,
\bar\psi_{-i}\sigma^i{}_j\psi_{+}{}^{j}-{\mit c.c.}$  This coupling
breaks the $U(k)_L\times U(k)_R$ chiral symmetry of the fermion
kinetic energy down to a diagonal $U(k)$.  Therefore, when we integrate
out the fermions to get an effective action for $\sigma$, we are dealing
with the standard problem of integrating out massive fermions that
receive their mass from spontaneous chiral symmetry breaking.  It is
precisely in connection with this problem that the anomalous Wess-Zumino
interaction, defined in \wzfun, was originally discovered.  The long
wavelength limit of the effective interaction obtained by integrating
out $\sigma$ is therefore -- allowing for the $N$ multiplets --
precisely
$$ L_{{\mit eff}}(\sigma)= N\Gamma(\sigma).   \eqn\mirro$$
The form of this interaction is completely determined by $U(k)_L\times
U(k)_R$ invariance and  the chiral anomaly.  (Let me warn that reader
that the $U(k)_L\times U(k)_R$ symmetry just invoked
is explicitly broken by interactions, such as the gauge couplings, that
do not contribute to $L_{{\mit eff}}(\sigma)$ in leading order in
$1/N$.  The corrections to the leading large $N$ behavior are the subject
of the next sub-section.)

Now we include the gauge fields and Feynman diagrams such as
that of figure (\fermloop).
Such diagrams must extend \mirro\ to a gauge invariant
effective action $L_{{\mit eff}}(\sigma,v)$.  The minimal choice,
in some sense, is the gauge invariant extension of the Wess-Zumino
action that was defined in \qqq:
$$L_{{\mit eff}}(\sigma,v)=N\Gamma(\sigma,v)     .  \eqn\riffo$$

Is this minimal form correct?
Apart from terms that vanish by the equations of motion
and terms of higher dimension that can be ignored at long distances,
a non-minimal gauge invariant term (on a flat world sheet)
would have to be of the form
$$\int_\Sigma  \Tr F W(\sigma), \eqn\rripo$$
with $F=dv+v\wedge v$ and $W$ a function of $\sigma$
that transforms in the adjoint representation.
In the next sub-section, we will show that
such a term is not generated, even by
corrections to the $1/N$ expansion.  We will also discuss the role
of the terms that vanish by the equations of motion, and some
curvature-dependent terms.

\subsection{Synthesis}

If we take the Lagrangian $N\Gamma(\sigma,v)$ by itself, it describes
a level $N$ gauged WZW model of $U(k)/U(k)$.  This sort of model
was analyzed in \S2, and as we know from \S2.5, it describes
a topological field theory.
There are no propagating modes at all, massless or massive.
If one adds conventional kinetic energy for $\sigma$ and $v$
(such terms are certainly present in our underlying Lagrangian),
one has propagating modes but massive ones.  Indeed the conventional
kinetic energy is irrelevant in the infrared and the large distance
behavior is that of the gauged WZW model.

Thus, the Grassmannian sigma model --
even if one does not restrict {\it a priori} to its topological sector --
reduces at long distances to a topological field theory.
In fact, any theory with a mass gap
will do this, since at distances at which the massive
particles can be neglected, all that survives is
dynamics of the vacuum or topological field theory.

In the case of the Grassmannian sigma
model, there was an underlying topological sector, described in \S4.2, and
visible from the classical Lagrangian before any analysis of its
quantum properties.  The basic observable in this topological sector
was the $\sigma$ field that appears in
\riffo\ (but now restricted to $\bar\sigma\sigma={\rm constant}$).
Thus the topological sector, defined microscopically, passes over
at large distances to the gauged WZW model governing the $\sigma$ field.
This is the mapping from the quantum cohomology of the Grassmannian
to the gauged WZW model (and thence the Verlinde algebra) that is the
main goal of this paper.

In what follows, we will analyze the
corrections to the $1/N$ expansion and eventually pin down the details
of the mapping from the quantum cohomology to the gauged WZW model.

\subsection{Search For Manifest Supersymmetry}

It would be attractive to find an extension of
\riffo, including the fermi partners of $\sigma$ and $v$, with
manifest $\N=2$ supersymmetry.  Of course, the one-loop effective
action from which \riffo\ was defined is such an extension, but
it would be nice to find, for instance, a compact description in $\N=2$
superspace of a local interaction describing
the long-wavelength part of the one-loop effective action.
I have been unable to do this and leave it as an interesting question.

However, let us truncate to
the abelian case in which $\sigma$, $v$, and their fermionic partners
are diagonal matrices (for instance
$\sigma={\rm diag}(\sigma_1,\dots,\sigma_k)$).
In this case, the field strength is similarly diagonal
(say $\Sigma={\rm diag}(\Sigma_1,\dots,\Sigma_k)$).
With this truncation,
it is possible to find an explicit, local
superspace interaction that describes
all of the anomalous interactions.  This interaction (which in the abelian
case was discussed in [\phases], \S3.2), is
$$L_{\N=2}={1\over \sqrt 2}\sum_{i=1}^k\int d^2x \,d\theta^+\,d\bar\theta^-
\left({it\Sigma_i\over 2}-{N\over 2\pi}\Sigma_i\ln(\Sigma_i/\mu)\right)+{\mit
c.c.}      \eqn\buffalo$$
The ease of writing
this interaction in the diagonal case and the difficulty of describing
its full non-abelian generalization may be related to the utility
of abelianization in the next subsection.

\section{Corrections}

Now we turn to analyzing the corrections
to this approximation.
We can ignore operators of dimension higher than two, which are irrelevant
at long distances.  Terms of dimension less than two, such as \nxon, cannot
arise as they would
violate the underlying $\N=2$ supersymmetry or (as a consequence) the
topological invariance of the twisted sector.
Also, we can ignore terms that vanish by the $v$
equations of motion and so can be eliminated, as in \S2.5, by a redefinition
of $v$.  Such terms (which are of the form $\int_\Sigma \sigma^*(B)$, with
$B$ an adjoint-invariant two-form on $U(k)$) would play no role in
our subsequent analysis.

We are left with three issues to consider:

(1) First, there might be corrections to the discrete data, the
``level'' of the effective
WZW model.  In the one loop approximation, we found the level to be
$N$; however, corrections of relative order $1/N$ could shift this by
a constant.
Actually, this has to be formulated more precisely because
the Lie algebra of $U(k)$, which we will call $u(k)$, is not simple;
it can be split as $su(k)\oplus u(1)$, where $su(k)$  consists of the
traceless $k\times k$ hermitian matrices and $u(1)$ is the center of $u(k)$.
In general, one could have a gauged WZW model for $U(k)$ of level
$(N_1,N_2)$, by which I mean that the Lagrangian would be determined
by the quadratic form $(\cdot,\cdot)$
on $u(k)$ such that $(a,b)=N_1\Tr ab$ for
$a,b\in su(k)$, and $(a,b)=N_2\Tr ab$ for $a,b\in u(1)$.
Thus, the first correction to the $1/N$ approximation might lead to
$(N_1,N_2)=(N+u,N+v)$ where $u,v$ are integers (perhaps depending on $k$);
higher order corrections in $1/N$ must vanish as
they could not be integral for all $N$.

(2) Second, the low energy effective action might contain a term
$$\int_\Sigma  \Tr F W(\sigma), \eqn\ripo$$
with as above $F$ the $u(k)$ curvature, and $W$ a function of $\sigma$
that transforms in the adjoint representation.  Though this term vanishes
by the equations of motion of the low-energy gauged WZW model and so could
be eliminated even from the quantum theory
by a field redefinition (as described in \S2.5),
it could still play a role that will be explained later.

Note that a constant term in $W$ (that is, a multiple of the identity)
could be absorbed in an additive
renormalization of $t$; we will not try to determine such a renormalization,
and all of our statements about $W$
will hold modulo an additive constant.

(3) Finally, we need to know whether, when the twisted theory is formulated
on a curved world-sheet, the effective Lagrangian contains a term
$$\Delta L = \int_\Sigma d^2x\sqrt h {R\over 4\pi} U(\sigma), \eqn\ocx$$
with $R$ the world-sheet curvature and $U(\sigma)$ a function invariant
under conjugation.  As we have discussed in \S2.5, any continuous
deformation of the gauged WZW model that preserves the topological
invariance and cannot be eliminated by a change of variables is of this
form.  (By contrast, the deformations considered above in (1) are discrete,
not continuous, and the deformations in (2) can be removed by a change of
variable.)

Now, here are the answers that I will claim for these three questions:

(A1) I will claim that the level of the effective gauged WZW model
is really $(N-k,N)$.  The correction can be thought of as a $1/N$
correction that comes from integrating out the $U(k)$ gauge multiplet
(the $u(1)$ level is not shifted since the gauge multiplet is neutral
under $u(1)$).

(A2) I will claim that $W=0$, in other words that no term of the form
\ripo\ is generated.

(A3) I will claim that a term of the form \ocx\ is generated, with
$$U = (N-k)\ln \det \sigma+{\rm constant}.\eqn\onzo$$
This might be regarded as the minimal possibility compatible with
the anomaly formula \oxoc.

\subsection{Abelianization}

Now I will explain how I will do the calculation.  A $1/N$ expansion
will not suffice, since we do not want to be limited to sufficiently
large $N$.  Instead, we will study the theory in the alternative
regime of $r<<0$.

To identify the corrections to the effective action of the three
types discussed above, it suffices to work in the region of field space
in which $\sigma$ is a diagonal matrix with distinct eigenvalues
$\sigma_1,\dots,\sigma_k$.  Moreover, we impose the condition
of unbroken supersymmetry (or vanishing vacuum energy); in the approximation
of \tarble\ -- which is valid for $r<<0$ -- the condition is
$$\bar\sigma_i\sigma_i =\mu^2
\exp(-4\pi r/N), ~{\rm for}~i=1,\dots,k.\eqn\urfo$$

The distinct values of the $\sigma_i$ break $U(k)$ to a diagonal subgroup
$U(1)^k$.  Calculations are relatively easy because the chiral superfields
and the ``off-diagonal'' part of the gauge multiplet
have large masses, of order $\bar\sigma\sigma$,
which can be read off from the classical Lagrangian.  The fields which
remain massless in this approximation (and actually get masses at one
loop, smaller by a factor of $e^2$) are the diagonal part of the gauge
multiplet.  The effective action for the massless modes, including the
one loop correction, has already been written with manifest $\N=2$
supersymmetry in \buffalo.   This is a kind of gauged WZW model
of $U(1)^k$.
So in this regime, we get a kind of abelianization of the Grassmannian
sigma model.

This should not come as a complete surprise, since as we have recalled
in \S2.6, the gauged WZW model has a precisely analogous abelianization.
Now, we will have to be careful in using abelianization to compute
the effects of types (1),
(2), and (3), because in going from the
gauged WZW model to its abelianization, precisely analogous terms
are generated.  These were computed in [\blau] and described in \S2.6,
and are as follows.

(B1) The shift in level in going from the
gauged WZW model of $U(k)$ to its abelianization is
$(k,0)$.  (There is obviously no shift of the $u(1)$ level under abelianization
since $u(1)$ is already abelian.)

(B2) No term of the form \ripo\ is generated.

(B3) The term of the form \ocx\ that is generated in abelianizing
the gauged WZW model was presented in equation \lateruse.

Now in verifying claims (A1), (A2), and (A3), we will integrate
out from the Grassmannian sigma model the
fields that, in the abelianized regime, have tree level masses;
thus we will get the precise abelianized
theory that is equivalent to the Grassmannian sigma model.
Then we will interpret the result as a sum of two contributions:
the terms claimed in (A1), (A2), and (A3) which describe how
to go from the topological sector of the Grassmannian sigma model
to an equivalent gauged WZW model; and the terms (B1), (B2), and (B3),
which arise in abelianization of the gauged WZW model.

So claims (A1), (A2), and (A3) are equivalent to the following claims,
which are the ones that we will actually check:

(C1) After abelianization, there is no shift in the level of the
Grassmannian sigma model from the naive result $(N,N)$.  We interpret
this to mean that the topological sector of the Grassmannian sigma
model is equivalent to a gauged WZW model of $U(k)/U(k)$
at level $(N-k,N)$, and the
level of that model is shifted by $(k,0)$ upon abelianization.

(C2) There will be no induced term of the type \ripo, in abelianizing
either of the two models or in comparing them.

(C3) The induced term of type \ocx\ in abelianization of the Grassmannian
model will be the sum of \onzo\ and the contribution \lateruse\ that arises
in abelianizing the gauged WZW model.
The sum of these is simply
$$\widetilde U(\sigma)=(N-1)\ln \det \sigma-\sum_{i\not=j}\ln(\sigma_i
-\sigma_j).       \eqn\oczo$$

\subsection{The Calculation}

Now I will explain the calculation justifying (C1), (C2), and (C3).
In discussing (C1) and (C2), world-sheet curvature is irrelevant, and we can
work on a flat ${\bf R}^2$.

(C1) and (C2) can be taken together and deduced from the following
principle.  Suppose that a $u(1)$ gauge field $v$, with field strength
$f=d v$,  is coupled
to a Dirac fermion $\chi$, of charge $q$. Let $\chi$ have a mass term
$$ L_{{\mit mass}}=-\int_\Sigma d^2x\left(\bar\chi_- m \chi_++\bar\chi_+
\bar m \chi_-\right). \eqn\gogog$$
We want to integrate out $\chi$ to get an effective action for $v$.
The dependence of the effective action on the phase of $m$ comes only
from the chiral anomaly and is
$$L_{{\rm eff}}= \dots +q\int_\Sigma  {f\over (2\pi)}{\rm Im }\log m.
\eqn\homer$$

Now we look at the Grassmannian sigma model in the abelianized regime
of $r<<0$, $\sigma_i$ large (obeying \urfo) and distinct.
The chiral superfields $\Phi^{js}$ and the off-diagonal part of the
gauge multiplet have bare masses of order $|\sigma_i|$.  They can be
integrated out in a one loop approximation; higher order corrections
would be of order $e^2$ and irrelevant.  Integrating out massive
bosons does not give terms relevant to (C1) or (C2), while
the contributions of fermions can be deduced from \homer.

To do so explicitly, let $v_i, ~i=1\dots k$, be the diagonal components
of the gauge field.   First we work out the contributions of chiral
superfields.  Each $v_i$ is coupled to $N$ chiral superfields $\Phi^{is},
{}~s=1\dots N$, of charge 1.  The fermi elements of these superfields
have mass $\sqrt 2\sigma_i$ (by inspection of \qqmmw), so their contribution
is
$$     N\sum_{i=1}^k\int_{\Sigma}{d v_i\over 2\pi}{\rm Im}\ln \sigma_i.
      \eqn\ity$$
This is the level $N$
gauged WZW action computed of equation
\riffo, specialized to the
case that only the diagonal components of $v$ and $\sigma$ are non-zero.

Now, we come to the off-diagonal part of the gauge fields.  Again,
the relevant contribution comes from the phases of the masses of the
off-diagonal fermions $\lambda^i{}_j,\,\,\,i\not=j$.
Since \ity\ coincides with the level $N$ gauged WZW action, the claims
(C1) and (C2) amount to the assertion
that no additional contribution will come from integrating out
the $\lambda^i{}_j$.

By inspection of \muccdo, the mass of $\lambda^i{}_j$ is $\sqrt 2(\sigma_i
-\sigma_j)$.  The gauge field $v_i$ interacts with the $\lambda^i{}_j$,
$j\not= i$, of charge 1, and with the $\lambda^m{}_i$, $m\not= i$,
of charge $-1$.  Their contribution adds up to
$$ \sum_{i,j}\int_\Sigma {dv_i\over 2\pi}\left({\rm Im}\ln(\sigma_i
-\sigma_j) -{\rm Im}\ln (\sigma_j-\sigma_i)\right). \eqn\jud$$
This is zero, or more exactly, it is independent of the $\sigma_i$.
Consequently it can be interpreted as a constant
term in $W$ or an additive renormalization of $t$; as noted in
the paragraph following \ripo, we will not keep track of such effects.
As for the diagonal components of $\lambda$, they are neutral and do not
couple to the $v_i$.

It remains to discuss (C3).  To this aim, we can take $\Sigma$ to
be a Riemann surface of genus zero and take the $\sigma_i$ to be constants.
As $\int_\Sigma d^2x \sqrt h R/4\pi=1$ in genus zero, the
claim (C3) amounts to the assertion that the
path integral $\int D\Phi_i\dots e^{-L}$ is
a constant multiple of
$$ (\det\sigma)^{-(N-1)}\prod_{i\not= j}(\sigma_i-\sigma_j). \eqn\huvvo$$

To verify this, we first integrate out the massive fields in the same
one loop approximation as above.  As before, the boson determinant
is real and depends only on $|\sigma_i|$,
while the fermion determinant has a phase that can be
extracted from the chiral anomaly.  Ordinarily, there is no chiral
anomaly for fermions in a gravitational field in two dimensions,
but the twisting to produce the topological theory involves
a modification of the fermion kinetic energy that introduces such an
anomaly.

We could proceed as above, starting with the anomaly formula analogous to
\homer.  For
the sake of variety, however, let us note that the anomaly can be
captured by the path integral over the zero modes of the fermion kinetic
energy.  For instance, the fermions $\psi^{is}$ from the
chiral multiplets have components $\bar\psi_+,\psi_-$ of spin zero
and other components of spin one.  The zero modes of the fermion
kinetic energy are the constant modes of $\bar\psi_+,\psi_-$, and
the path integral over those modes is
$$\int d\bar\psi_{+is}d\psi_-{}^{jt}\exp\left(\sum_{is}\bar\psi_{+is}
\bar\sigma_i\psi_-{}^{is}\right)=\det\bar\sigma^N={\rm constant}\cdot
\det\sigma^{-N}, \eqn\nuffo$$
where we have used the fact that $\sigma\bar\sigma = {\rm constant}$.
Similarly, for the off-diagonal $\lambda$ fields, the
zero modes of the kinetic energy are the constant modes of $\bar\lambda_-$,
$\lambda_+$, and the path integral over those modes is
$$\prod_{i\not = j}\int d\bar\lambda_-{}^i_j \,d\lambda_+{}^j{}_i
\exp\left(\sqrt 2
(\sigma_i-\sigma_j)\bar\lambda_-{}^i{}_j\lambda_+{}^j_i\right)
={\rm const}\cdot \prod_{i\not= j}\left(\sigma_i-\sigma_j\right).  \eqn\upplo$$

Comparing \nuffo\ and
\upplo\ with the claim made in \oczo\ concerning (C3), we
see that we are missing precisely one factor of $\det\sigma$.
This must come from the remaining integral over the diagonal
components of the gauge field.  Indeed, though the diagonal fermions
$\lambda^i{}_i$ are massless at tree level, they receive at the one
loop level a mass term with the form
$$ \sum_{i=1}^k\left(\lambda_+{}^i{}_i
\bar\sigma_i{}^{-1}\bar\lambda_-{}^i{}_i+{\mit c.c.}\right)
\eqn\mippo$$
This term can be straightforwardly calculated or can be read off from
the $\N=2$ extension \buffalo\ of the bosonic anomalous interactions.
Integrating out the diagonal fermions therefore gives
(up to a constant) a factor of
$$\left(\det\bar\sigma\right)^{-1}
 ={\rm constant}\cdot \left(\det\sigma\right).       \eqn\ovvo$$
This is the last factor needed for (C3).

This factor could in a more general way be predicted as follows.
Just because the diagonal theory (including the $\lambda^i{}_i$)
is a product of $k$ sub-theories,
the phase it produces must be of the form
$$\prod_{i=1}^k F(\sigma_i) \eqn\poxp$$
for some function $F$.  Given this factorized form, to agree
with the anomaly formula \oxoc\ it must be that $F(\sigma)=\sigma$.

\section{The Verlinde Algebra And The Grassmannian}

In this subsection, we will put the pieces together and write
down the precise connection between the Verlinde algebra and
the quantum cohomology of the Grassmannian.

First of all, we consider the map from the cohomology ring of the
Grassmannian to the Verlinde algebra.  The quantum cohomology ring
of $G(k,N)$ is generated by operators of the form
$$O_V=\Tr_V\sigma, \eqn\olfo$$
with $V$ an irreducible representation of $U(k)$.  Thinking of $\sigma$
is the curvature of the natural connection on the tautological rank
$k$ bundle over $G(k,N)$, $\Tr_V\sigma$ can be interpreted as a characteristic
class of that bundle and hence a cohomology class of $G(k,N)$.

On the other hand, working at long distances, $\sigma$ becomes
a unitary matrix (up to a constant that we will eventually pin down)
and then $\Tr_V\sigma$
can be interpreted as an operator in the effective gauged WZW model
of $U(k)/U(k)$.
We worked out in \finalgo,\josos\ the interpretation of this operator:
it is the element of the Verlinde algebra determined by the representation
$V$.  Of course, from what we have said in \S4.6, the Verlinde
algebra in question is the one for the group $U(k)$ at
level $(N-k,N)$.

So we have gotten the precise mapping from the cohomology of the
Grassmannian to the Verlinde algebra.  A couple of points should be
clarified:

(1)  It is essential that in mapping from the Grassmannian sigma
model to the gauged WZW model, there is no correction of type \ripo.
Such a term, since it vanishes by the equations of motion
in the gauged WZW model, could be transformed away by a redefinition
of $\sigma$ and $v$.  But the resulting redefinition of $\sigma$
would cause the operator $O_V$ to mix with similar operators for
other representations.  Thus, were terms with the structure \ripo\ to appear,
their precise form would enter in determining the map from
the cohomology of $G(k,N)$ to the Verlinde algebra.

By contrast, corrections to the gauged WZW action
that vanish by the $v$ equations of motion and so can be removed
by redefinition of $v$ are immaterial, since $O_V(\sigma)$ is
independent of $v$.  As noted at the beginning of \S4.3,
we have made no claim that corrections that vanish by the $v$
equations of motion are not generated or have any particular structure.

(2) In the rest of this paper, we will set the Kahler parameter $r$ to
zero; as explained in \S4.3, this involves no essential loss of information.
Two other constants discussed in \S4.3 also enter.
One is the constant $c$ in the relation
$\sigma = c\cdot {\rm unitary~matrix}$, which we evaluated only approximately
in \ommo.
The other is the additive renormalization constant called $a$ in \normert,
which is unknown since
we did not attempt to determine the constant in \onzo.
For the time being, we will set $c$ and $a$ to 1 and 0; eventually
we will verify that this is correct (for $r=0$)
by checking special cases of the formulas.

\subsection{Correlation Functions And The Metric}

Now, let us determine precisely how the correlation functions
and the metric in the Grassmannian sigma model compare to those
in the gauged WZW model.  The essential point that goes beyond
what we have just said above is that one must include the correction
term of \ocx, \onzo.  By topological invariance, $\sigma$
can be treated as a constant, so the correction factor in the path
integral is
$$\exp(-\Delta L)=\exp\left(-\int_\Sigma d^2x\sqrt h{R\over 4\pi}
(N-k)\ln\det\sigma\right) = (\det\sigma)^{(g-1)(N-k)}.  \eqn\bumbo$$
The point or points at which $\det\sigma$ is inserted are immaterial.
(It will turn out that $\det \sigma$ is an invertible element of the
Verlinde algebra or quantum cohomology.)

So if $\langle ~~~\rangle_{G(k,N)}$ denotes an expectation value of
the path integral of the Grassmannian sigma model, and $\langle ~~~\rangle
_{{\mit WZW}}$ denotes a path integral in the gauged WZW model,
then the relation between these symbols in genus $g$
is
$$\left\langle \prod_{i=1}^s \Tr_{V_i}(\sigma_i)\right\rangle_{G(k,N)}
=\left\langle\prod_{i=1}^s \Tr_{V_i}(\sigma_i)  \cdot (\det\sigma)^{(g-1)
(N-k)}\right\rangle_{{\mit WZW}}. \eqn\omigo$$
Henceforth we will abbreviate $\Tr_{V_i}(\sigma_i)$ as $V_i$.
{}From \omigo\ one can see that the natural metric on the cohomology
of $G(k,N)$ (given by Poincar\'e duality) does not coincide with
the natural metric on the Verlinde algebra.  Let us call these
metrics (the sigma model and Verlinde metrics)
$g_{{\sigma}}$ and $g_{{V}}$, respectively.
We recall that the metric is defined by a two point function in
genus 0, so
$$\eqalign{  g_{{\sigma}}(V_1,V_2) & = \left\langle V_1V_2
                                \right\rangle_{G(k,n)}\cr
             g_{{V}}(V_1,V_2)& =\left\langle V_1V_2\right\rangle_{{\mit WZW}}=
\langle V_1V_2\cdot
(\det\sigma)^{(N-k)}\rangle_{G(k,N)},\cr} \eqn\controv$$
with the correlation functions being in genus zero.

Now let us compare the ring structure on the cohomology of the
Grassmannian to the ring structure of the gauged WZW model.
We recall that in either of the two theories,
the ring structure is introduced by interpreting
the genus zero three point function in terms of a binary operation,
say $V_1,V_2\to V_1\cdot V_2$, according to the following formula:
$$\left\langle V_1V_2V_3\right\rangle =g(V_1\cdot V_2,V_3). \eqn\murmoro$$
The relation between the genus zero three point functions of the
two theories is from \omigo\
$$\left\langle V_1V_2V_3\right\rangle_{{\mit WZW}}=\langle V_1V_2V_3
(\det\sigma)^{(N-k)}\rangle_{{\mit G(k,N)}}. \eqn\urmoro$$
In particular the three point functions of the $G(k,N)$ and
gauged WZW models do not coincide.  However, they differ by the
same factor of $\left(\det\sigma\right)^{(N-k)}$ that
enters in the relation between the metrics.
This means in fact, upon putting together the last few formulas,
that the multiplication laws are the same in the two theories.

So finally, our natural map from the quantum cohomology of the Grassmannian
to the Verlinde algebra  is a ring homomorphism -- justifying terminology
that was used above.

\subsection{Non-Abelian Theta Functions}

In the title of this paper and in much of the writing of it,
I have emphasized the Verlinde algebra, which determines the
dimension of the space of non-abelian theta functions.  However,
the above gives directly a formula for the dimension of the space
${\cal H}$ of non-abelian theta functions without having to detour via the
Verlinde algebra.  Of course, we will count non-abelian theta
functions for the group $U(k)$ at level $(N-k,N)$; because the
$U(1)$ theory is well understood, there is no essential difficulty
in generalizing to other levels.

Let $\langle 1\rangle^g$ denote the partition function in genus $g$.
Then the dimension of ${\cal H} $ on
a Riemann surface $\Sigma$ of genus $g$ is
$$\dim{\cal H}=\langle 1\rangle^g_{{\mit WZW}}
=\left\langle (\det\sigma)^{-(g-1)(N-k)}\right\rangle_{G(k,N)}
=\left  \langle(\det\sigma)^{k(g-1)}\right\rangle_{G(k,N)}
\eqn\yutt$$
In the last expression, I have used the fact that $(\det\sigma)^N=1$,
as one can deduce from the Landau-Ginzburg description
of the quantum cohomology (we do this below for $k=2$).
The right hand side of \yutt\
can be evaluated by counting holomorphic maps of $\Sigma$ to
$G(k,N)$ obeying certain conditions.

\section{Getting Down To Earth}

In this section, we will make everything completely explicit
in the cases of $k=1$ and $k=2$.  (Some of the issues are discussed
by Gepner for general $k$ in the last paper in [\gepner].)

First we dispose of $G(1,N)$, that is, ${\bf CP}^{N-1}$.
Over ${\bf CP}^{N-1}$ there is a tautological principal $U(1)$ bundle
$P$.  Let $W$ be the standard representation of $U(1)$, of
``charge one.''  Associated to $P$ in the representation $W$ is
a line bundle ${\cal L}$. The representation $W$ determines
an operator $\Tr_W\sigma=\sigma$ which we will call $x$.
In the sigma model, interpreting $\sigma$ as the curvature of the natural
connection on ${\cal L}$, $x=c_1({\cal L})$.

The cohomology ring of ${\bf CP}^{N-1}$ is generated
by $x$ and classically is ${\bf C}[x]/x^{N-1}$.  But
from [\yaulect] or the $k=1$ case of \ellform, the quantum cohomology
ring is
$$ R={\bf C}[x]/(x^N-1).           \eqn\urmo$$
The metric on the cohomology determined by Poincar\'e duality is
$$g_{{\sigma}}(x^k,x^l) =\delta_{k+l,N-1}, \eqn\nurmo$$
where in view of \urmo,
$k$ and $l$ are evaluated modulo $N$.

On the other hand, in the gauged WZW model, $x=\Tr_W\sigma$
should be identified with the element of the Verlinde algebra
for $U(1)$ at level $N$ determined by the representation $W$.
The structure of this algebra is well known.
It is generated by $W$ with the relation $W^N=1$, just
as in \urmo, and the metric is
$$g_{{V}}(W^k,W^l)=\delta_{k+l,0} ,\eqn\rmo$$
with again $k$ and $l$ taken modulo $N$.
\nurmo\ and \urmo\ are related in the fashion predicted by \controv.
Indeed since for either metric $g(a,b)=g(ab,1)$, \controv\ is equivalent
to
$$g_{{V}}(x^k,1)= g_{{\sigma}}(x^k,x^{N-1}).  \eqn\udlo$$

That disposes of $k=1$.  Obviously, we cannot expect the non-abelian
case $k=2$ to be as trivial as that.

\subsection{The Verlinde Algebra For $k=2$}

First we describe explicitly (but not in a fully self-contained
fashion) the Verlinde algebra of the group
$U(2)$ at the desired level $(N-2,N)$.

We have an exact sequence
$$ 1\to Z_2\to SU(2)\times U(1)\underarrow{f} U(2)\to 1.\eqn\ippo$$
Here the map $f$ is as follows: we identify $SU(2)$ with the $2\times 2$
unitary matrices of determinant 1, $U(1)$ with $2\times 2$ unitary
matrices that are multiples of the identity, and for $x\in SU(2)$,
$y\in U(1)$, let $f(x,y)=xy$.

The gauged WZW action of $U(2)$ at level $(N-2,N)$ restricts,
if one takes the fields to lie in $SU(2)$, to the $SU(2)$ action
at level $N-2$; but if one takes the fields to be in $U(1)$, it restricts
to the $U(1)$ gauged WZW action at level $2N$.  (A factor of two arises
simply because the trace of the identity matrix in the fundamental
representation of $U(2)$ is 2.)  Therefore, we will proceed
by comparing the Verlinde algebra of $U(2)$ at level $(N-2,N)$
to that of $SU(2)\times U(1)$ at level $(N-2,2N)$.

The $SU(2)$ Verlinde algebra was described explicitly in [\bott].
If $V_1$ is the two dimensional representation of $SU(2)$,
and $V_n$ is its $n^{th}$ symmetric tensor power, then the Verlinde
algebra of $SU(2)$ is the usual representation ring of $SU(2)$,
subject to the relation
$$V_{N-1}=0. \eqn\subrel$$
The representation ring of $SU(2)$, subject to this relation,
is spanned additively by $V_0,\dots,V_{N-2}$.
The multiplication law can be described explicitly as
$$V_i\times V_j=\sum_t N_{ijt}V_t, \eqn\jroo$$
where $N_{ijt} $ is 1 if the following relations and their
cyclic permutations are obeyed:
$$i+j\geq t, ~~ N-2-i+j\geq N-2-t,~~ 2(N-2)-i-j\geq t.\eqn\hugf$$
Otherwise $N_{ijt}=0$.
The metric on the Verlinde algebra is
$$g(V_s,V_t)=\delta_{s,t}. \eqn\ubrel$$

The $U(1)$ Verlinde algebra has already been introduced above.
It is generated by the charge one representation $W$, and
the defining relation at level $2N$ is
$$W^{2N}=1. \eqn\hubrel$$
The metric is
$$g(W^u,W^v)=\delta_{u+v,0}. \eqn\bgrel$$

The $SU(2)\times U(1)$ Verlinde algebra at level $(N-2,2N)$,
is therefore spanned additively by the elements $V_iW^j$,
for $i=0,\dots,N-2$, $j=0,\dots,2N-1$, corresponding
to the representation $V_i\otimes W^{\otimes j}$.
The multiplication law
and metric are products of the multiplication law and metric
of $SU(2)$ and $U(1)$.

Now we want to proceed to $U(2)=(SU(2)\times U(1))/{\bf Z}_2$.  Dividing
by ${\bf Z}_2$ halves the volume of the group manifold (if one
uses a fixed Haar measure in an obvious sense).  The Verlinde algebra
is defined on a certain space of conformal blocks which can be constructed
by quantizing an appropriate phase space ${\cal M}$
-- for instance, the phase
space of the gauged WZW model.  When the volume of the
group is halved, the volume of ${\cal M}$ is divided by $2^2=4$\foot{
${\cal M}$ is the moduli space of flat connections in genus one and
consists of pairs of commuting elements of the gauge group $G$, divided by
the Weyl group.  Because one has a pair of elements of $G$,
the volume of ${\cal M}$ is decreased by a factor of $n^2$ if one
divides $G$ by a group of order $n$.};
therefore, in the semiclassical limit of large $N$, the Verlinde algebra
of $U(2)$ at level $(N-2,N)$ will have one fourth the dimension of
that of $SU(2)\times U(1)$ at level $(N-2,2N)$.

\REF\zoo{G. Moore and N. Seiberg, ``Taming The Conformal Zoo,''
Phys. Lett. {\bf B220} (1989) 422.}
One factor of two is obvious.  Among all $SU(2)\times U(1)$ representations,
we must restrict to those that are representations of $U(2)$.
This means keeping only $V_iW^j$ with $i+j$ even.
The second factor of two is less obvious.
One must impose the equivalence
relation
$$ V_iW^j=V_{N-2-i}W^{j+N}. \eqn\grort$$
I refer the interested reader to [\zoo] for an explanation
(in the analogous case of $SO(3)=SU(2)/{\bf Z_2}$) of such matters.

Note that if we set $\tau(V_iW^j)=V_{N-2-i}W^{j+N}$,
then the Verlinde algebra of $SU(2)\times U(1)$ at level
$(N-2,N)$ obeys $\tau(a)b=a\tau(b)=\tau(ab)$.
This ensures that the Verlinde algebra of $SU(2)\times U(1)$
induces a natural algebra structure on the quotient by the relations \grort.
This quotient algebra, restricted to $i+j$ even (a $\tau$-invariant
condition) is the Verlinde algebra of $U(2)$ at level $(N-2,N)$.
The metric is
$$g_{{V}}(V_iW^s,V_jW^t)=\delta_{i,j}\delta_{s+t,0}
+\delta_{i,N-2-j}\delta_{s+t-N,0}.  \eqn\ugglo$$

A complete but redundant set of relations for the $U(2)$ Verlinde
algebra would be the relations
$$ V_{N-1}=0,~~~W^{2N}=1  \eqn\gglo$$
inherited from the $SU(2)$ and $U(1)$ algebras, along with \grort.
A special case of \grort\ is
$$ V_{N-2}W^N = 1.   \eqn\gloff$$
It will become clear presently that \gglo\ and \gloff\ suffice to
characterize the Verlinde algebra.

\subsection{Representations And Characters}

Two representations of the group $U(2)$ will play
a distinguished role.  The first is the standard two dimensional
representation ${\cal V}_1$.  Under restriction to $SU(2)\subset U(2)$,
${\cal V}_1$ restricts to the standard two dimensional representation
$V_1$ of $SU(2)$, and the scalars in $U(2)$ act with charge 1.
So ${\cal V}_1$ pulls back to
the representation $V_1\otimes W$ of $SU(2)\times U(1)$.

The other important representation is $\eta=\wedge^2{\cal V}_1$.
$SU(2)$ acts trivially on $\eta$, and the scalars in $U(2)$ act
with charge 2, so $\eta$ pulls back to the representation $W^2$
of $SU(2)\times U(1)$.

So the elements of the $U(2)$ Verlinde algebra determined
by ${\cal V}_1$ and $\eta$ are just $V_1W$ and $W^2$.
What elements in the quantum cohomology do these same representations
determine?  Over $G(2,N)$, there is a tautological complex two-plane
bundle $E^*$ with curvature matrix represented by the quantum field
$\sigma$.  The Chern classes of $E^*$ are
$$\eqalign{ c_1(E^*) & = \Tr_{{\cal V}_1}\sigma \cr
            c_2(E^*) & = \Tr_\eta\sigma . \cr}
\eqn\porfori$$
So under the natural mapping from the quantum cohomology of $G(2,N)$
to the Verlinde algebra of $U(2)$, $c_1(E^*)$ and $c_2(E^*)$ correspond
to the representations ${\cal V}_1$ and $\eta$.

Let us now briefly discuss the classical representation ring of $U(2)$;
the Verlinde algebra is a quotient of this, as we have described.
Consider the maximal abelian subgroup of $U(2)$ of matrices
of the form
$$\sigma=\left(\matrix{ \lambda_1& 0 \cr 0 & \lambda_2\cr }
    \right),  \eqn\polon$$
with $|\lambda_1|=|\lambda_2|=1$.
The character of a representation $R$ of $U(2)$ is $\Tr_R\sigma$
regarded as a function of the $\lambda_i$.
For instance, the characters of ${\cal V}_1$ and $\eta$
are
$$\eqalign{ \Tr_{{\cal V}_1}\sigma & = \lambda_1+\lambda_2 \cr
            \Tr_\eta\sigma & = \lambda_1\lambda_2.  \cr}\eqn\holon$$
If ${\cal V}_n$ is the $n^{th}$ symmetric tensor power of ${\cal V}_1$,
then its character is
$$\Tr_{{\cal V}_n}\sigma ={\lambda_1{}^{n+1}-\lambda_2{}^{n+1}\over
\lambda_1-\lambda_2}.       \eqn\bolon$$
Any irreducible representation of $U(2)$ is of the form ${\cal V}_s\eta^t$ for
some integers $s,t$ (with $s\geq 0$).  The equivalence relation
\grort\ becomes
$${\cal V}_s\eta^t\leftrightarrow {\cal V}_{N-2-s}\eta^{s+t+1}.\eqn\newgrort$$

In general, the map from a representation of $U(2)$ to its character
is an isomorphism between the ring of representations of $U(2)$
and the ring of Laurent polynomials in $\lambda_1$ and $\lambda_2$
that are invariant under the Weyl group, which acts by $\lambda_1
\leftrightarrow\lambda_2$.

As we have sketched above, the Verlinde algebra of $U(2)$ is a quotient
of the classical representation ring of $U(2)$ by a certain ideal.
Under the isomorphism  between the representation ring and the
character ring, the  generators of this ideal can be identified
with certain Laurent polynomials in the $\lambda$'s.  For instance,
the relations in \gglo\ and \gloff\ become
$$\eqalign{{\lambda_1{}^N-\lambda_2{}^N\over\lambda_1-\lambda_2} & = 0 \cr
            (\lambda_1\lambda_2)^N & = 1  \cr
            \lambda_1\lambda_2{\lambda_1{}^{N-1}-\lambda_2{}^{N-1}
      \over\lambda_1-\lambda_2} & = 1 .\cr   }  \eqn\hiddo$$

The second relation in \hiddo\ has the following implication.  The
classical representation ring of $U(2)$ is a ring of {\it Laurent}
polynomials in the $\lambda$'s, including negative powers.  But by
multiplying by a suitable power of $1=(\lambda_1\lambda_2)^N$,
one can clear the denominators and regard the Verlinde algebra as
the quotient of the ring of Weyl-invariant polynomials (not Laurent
polynomials)  in the $\lambda$'s
by a certain ideal ${\cal I}$.  We will learn that ${\cal I}$
is in fact generated
by the first and third relations in \hiddo.

If we multiply the first relation in \hiddo\ by $\lambda_1+\lambda_2$
and subtract the third,
we learn that
$${\lambda_1{}^{N+1}-\lambda_2{}^{N+1}\over \lambda_1-\lambda_2
}+1 = 0.    \eqn\triddo$$

\subsection{Cohomology Ring Of $G(2,N)$}

Now return to the sigma model interpretation of $\sigma$ as the curvature
of the tautological two-plane bundle $E$ over $G(2,N)$.
If we introduce the roots $\widetilde \lambda_1,\widetilde
\lambda_2$ of the Chern polynomial,
then the Chern classes of $E$ are
$$\eqalign{ c_1(E^*) & = \widetilde \lambda_1+\widetilde\lambda_2\cr
            c_2(E^*) & = \widetilde \lambda_1\widetilde
\lambda_2 .\cr}  \eqn\uncoc$$
We observed earlier that
under the map from cohomology of $G(2,N)$ to the Verlinde algebra,
$c_1(E^*)$ and $c_2(E^*)$ correspond to ${\cal V}_1$ and $\eta$.
If in turn we identify the Verlinde algebra as a quotient of the character
ring, ${\cal V}_1$ and $\eta$ are identified with their characters, which
were given in \holon.  As \holon\ and \uncoc\ coincide,
the identification between these rings can be interpreted as
$\lambda_i\leftrightarrow\widetilde\lambda_i$.  Henceforth, we make
this identification and drop the tildes.

The quantum comology ring of $G(2,N)$ is the ring of polynomials in
$c_1(E^*)$ and $c_2(E^*)$ modulo an ideal ${\cal J}$.  As explained in
\S3.2, ${\cal J}$ can be described as follows.   Let
$$W(\lambda_1,\lambda_2) ={1\over N+1}\left(\lambda_1{}^{N+1}
+\lambda_2{}^{N+1}\right)+\left(\lambda_1+\lambda_2\right).  \eqn\cncncn$$
Since it is Weyl-invariant, $W$ can be regarded as a polynomial
in $c_1=\lambda_1+\lambda_2$
and $c_2=\lambda_1\lambda_2$.
The ideal ${\cal J}$ is generated by the relations
$$0=dW= {\partial W\over\partial c_1}dc_1+{\partial W\over\partial c_2}dc_2.
                \eqn\brno$$

Since
$$\eqalign{ d\lambda_1=&{\lambda_1 dc_1-dc_2\over \lambda_1-\lambda_2}\cr
            d\lambda_2=&{-\lambda_2dc_1+dc_2\over\lambda_1-\lambda_2},\cr
}\eqn\cxzs$$
we have
$$dW=\left(\lambda_1{}^N+1\right)d\lambda_1+\left(\lambda_2{}^N+1\right)
d\lambda_2=dc_1\left({\lambda_1{}^{N+1}-\lambda_2{}^{N+1}\over
\lambda_1-\lambda_2}+1\right)-dc_2\left({\lambda_1{}^N-\lambda_2{}^N
\over\lambda_1-\lambda_2}\right).\eqn\polnson$$
The quantum cohomology of $G(2,N)$ is therefore defined by the relations
$$0={\lambda_1{}^N-\lambda_2{}^N\over \lambda_1-\lambda_2}
={\lambda_1{}^{N+1}-\lambda_2{}^{N+1}\over \lambda_1-\lambda_2}+1.\eqn\cract$$
If we compare this to \triddo\ and to the first equation in \hiddo,
we see that these relations hold in the Verlinde algebra,
and consequently the Verlinde algebra is a quotient of the
quantum cohomology of $G(2,N)$.

To show that the two algebras coincide (and that all additional
relations we found earlier for the Verlinde algebra are consequences
of \polnson), it suffices to compare the dimensions of the two algebras.
{}From the description of the Verlinde algebra as being
spanned by the elements $V_iW^j,$ with $0\leq i\leq N-2$, $0\leq
j\leq 2N-1$, with a two-fold restriction and a two-fold identification,
its dimension is $N(N-1)/2$.  On the other hand, with $c_1$ and
$c_2$ considered to be of degree 1 and 2, respectively,
the potential $W(c_1,c_2)$ is homogeneous of degree $N+1$; it follows
by a simple counting that the polynomial ring in the $c_j$ modulo
the ideal $dW=0$ has dimension $N(N-1)/2$.  This completes the
explicit verification of the equivalence between these rings.

Moreover, we can now dispose of the constant $c$ in the relation
$\sigma = c g$.  This constant corresponds to a possible constant
in the relation $\lambda_i\leftrightarrow\widetilde \lambda_i$.
The relations above such as \cract\ are not invariant
under rescaling of the $\lambda$'s, and so the agreement with the Verlinde
algebra would be ruined if we modified the value of $c$.  A similar
argument holds for $k>2$.

\subsection{The Metric}

It remains to show that the metric on the Verlinde algebra and the
metric on the quantum cohomology of $G(2,N)$ are related in the
expected fashion.

Since either metric obeys $g(a,b)=g(ab,1)$, to verify \controv,
it suffices to show that
$$g_{{
V}}(a,1)=g_{{\sigma}}(a,(\det\sigma)^{N-2}).\eqn\occovo$$
We already know the Verlinde metric:
$$g_{{V}}
({\cal V}_s\eta^t,1)=\delta_{s,0}\delta_{t,0} +\delta_{N-2-s,0}
\delta_{t-1,0}.\eqn\orfo$$

Now let us compute the metric on the cohomology of $G(2,N)$.  Since
there are no instanton corrections to the metric, we need only compute
the classical metric on the cohomology.
According to \jupper, that metric is
$$g_{{\sigma}}(a,b)=
-{1\over 2} \sum_{dW(\lambda)=0}
{ab(\lambda_1-\lambda_2)^2\over N^2\lambda_1{}^{N-1}\lambda_2{}
^{N-1}}=
-{1\over 2}\sum_{\lambda_1{}^N=\lambda_2{}^N=-1}
{ab(\lambda_1-\lambda_2)^2\over N^2\lambda_1{}^{N-1}\lambda_2{}^{N-1}}.
\eqn\nugl$$

The symmetric polynomials in the $\lambda$'s of degree
$2(N-2)$ (corresponding to the top dimensional cohomology of $G(2,N)$)
are of the form
$$ f_r = {\lambda_1{}^{2r+1}-\lambda_2{}^{2r+1}\over\lambda_1-\lambda_2}
(\lambda_1\lambda_2)^{N-2-r},\,\,\,{\rm with}~0\leq r\leq N-2.\eqn\truv$$
A simple calculation gives
$$g_{{\sigma}}(f_r,1)=-{1\over 2N^2}\sum_{\lambda_1{}^N=\lambda_2{}^N=-1}
\left(\left(\lambda_1\over\lambda_2\right)^{1+r}
-\left(\lambda_1\over\lambda_2\right)^{r}
-\left(\lambda_1\over\lambda_2\right)^{-r}
+\left(\lambda_1\over\lambda_2\right)^{-1-r}\right)=\delta_{r,0}.
\eqn\ilmoxx$$

This reproduces the first term on the right of \orfo\ up to the shift
predicted in \occovo.  To interpret the second term on the right of \orfo, note
that while classically for homogeneous $f$,
$g_{{\sigma}}(f,1)$ is non-zero unless $f$ is of degree $2(N-2)$,
the quantum cohomology is only graded modulo $N$ (in complex dimension),
so we can also consider the case that $f$ is of degree $N-4$;
by a calculation similar to the above, this reproduces the second term
in \orfo.

Moreover, we can now dispose of the renormalization constant $a$
of equation  \normert.  Inclusion of this term would rescale the metric
by a factor of $e^{2a}$; the agreement between the two metrics means
that the above formulas are normalized correctly, at $r=0$.
Though we have made this check on the value of $a$ (and a similar,
earlier check for $c$) only for $k=2$, the arguments are similar for any
$k$.

\ack{I benefited from discussions with G. Segal at an early stage
of this work.}
\refout
\figout
\end